\documentclass[aps,prx,twocolumn,showpacs,superscriptaddress,groupedaddress,floatfix]{revtex4-1}

\usepackage{graphicx}  
\usepackage{dcolumn}   
\usepackage{bm}        
\usepackage{amssymb}   
\usepackage{amsmath}   

\usepackage{color} 

\usepackage{multirow}
\usepackage{enumerate}
\usepackage{comment}
\usepackage{tabularx}
\usepackage{booktabs}

\usepackage{gensymb}
\usepackage{siunitx}

\newcommand{\ii}{{\rm i}}
\newcommand{\dd}{{\rm d}}

\def\lapp{\ifmmode\stackrel{<}{_{\sim}}\else$\stackrel{<}{_{\sim}}$\fi}
\def\gapp{\ifmmode\stackrel{>}{_{\sim}}\else$\stackrel{<}{_{\sim}}$\fi}

\begin{document}

\title{Lunar Gravitational-Wave Antenna}

\author{Jan Harms$^{1,2}$, Filippo Ambrosino$^{3,4,5}$, Lorella Angelini$^6$, Valentina Braito$^7$, Marica Branchesi$^{1,2}$, Enzo Brocato$^{8,9}$, Enrico Cappellaro$^{10}$, Eugenio Coccia$^{1,2}$, Michael Coughlin$^{11}$, Roberto Della Ceca$^{7}$, Massimo Della Valle$^{12}$, Cesare Dionisio$^{13}$, Costanzo Federico$^{4}$, Michelangelo Formisano$^{4}$, Alessandro Frigeri$^{4}$, Aniello Grado$^{12,14}$, Luca Izzo$^{15}$, Augusto Marcelli$^{16,17,18}$, Andrea Maselli$^{5,19}$, Marco Olivieri$^{20}$, Claudio Pernechele$^{10}$, Andrea Possenti$^{21}$, Samuele Ronchini$^{1,2}$, Roberto Serafinelli$^{7}$, Paola Severgnini$^{7}$, Maila Agostini$^{22,23}$, Francesca Badaracco$^{1,2}$, Lorenzo Betti$^{22,23}$, Marta Maria Civitani$^{7}$, Christophe Collette$^{24,25}$, Stefano Covino$^{7}$, Simone Dall'Osso$^{1,2}$, Paolo D'Avanzo$^{7}$, Matteo Di Giovanni$^{1,2}$, Mauro Focardi$^{26}$, Carlo Giunchi$^{27}$, Joris van Heijningen$^{28}$, Nandita Khetan$^{1,2}$, Daniele Melini$^{29}$, Giuseppe Mitri$^{30,31}$, Conor Mow-Lowry$^{32}$, Luca Naponiello$^{22,23}$, Vladimiro Noce$^{22,23}$, Gor Oganesyan$^{1,2}$, Emanuele Pace$^{22,23}$, Ho Jung Paik$^{33}$, Alessandro Pajewski$^1$, Eliana Palazzi$^{34}$, Marco Pallavicini$^{35,36}$, Giovanni Pareschi$^{7}$, Ashish Sharma$^{1,2}$, Giorgio Spada$^{37}$, Ruggero Stanga$^{22}$, Gianpiero Tagliaferri$^{7}$}

\affiliation{$^1$Gran Sasso Science Institute (GSSI), I-67100 L’Aquila, Italy}
\affiliation{$^2$INFN, Laboratori Nazionali del Gran Sasso, I-67100 Assergi, Italy}
\affiliation{$^3$INAF - Osservatorio Astronomico di Roma, Via Frascati 33, I-00078, Monte Porzio Catone (Rome), Italy} 
\affiliation{$^4$INAF - Istituto di Astrofisica e Planetologia Spaziali, Via Fosso del Cavaliere 100, I-00133, Rome, Italy}
\affiliation{$^5$Sapienza Universit\`a di Roma, Piazzale Aldo Moro 5, I-00185, Rome, Italy}
\affiliation{$^6$Astroparticle Physics,  NASA Goddard Space Flight Center , Greenbelt , MD 20771 USA} 
\affiliation{$^7$INAF - Osservatorio Astronomico di Brera, Via Brera 21, I-20121, Milan, Italy} 
\affiliation{$^8$INAF - Osservatorio Astronomico d'Abruzzo, Via M. Maggini snc, I-64100, Teramo, Italy}
\affiliation{$^9$INAF - Osservatorio Astronomico di Roma, Via Frascati 33, I-00078, Monte Porzio Catone (Rome), Italy}
\affiliation{$^{10}$INAF - Osservatorio Astronomico di Padova, Vicolo dell'Osservatorio 5, I-35122, Padova, Italy}  
\affiliation{$^{11}$School of Physics and Astronomy, University of Minnesota, Minneapolis, Minnesota 55455, USA} 
\affiliation{$^{12}$INAF - Osservatorio Astronomico di Capodimonte, Salita Moiariello 16, I-80131, Naples, Italy} 
\affiliation{$^{13}$Space Boy Station srl, Via Aldo Sandulli 45, I-00156, Rome, Italy} 
\affiliation{$^{14}$INFN - Sezione di Napoli, Via Cintia, I-80126, Naples, Italy} 
\affiliation{$^{15}$DARK, Niels Bohr institute, University of Copenhagen, Jagtvej 128, DK-2200 Copenhagen \O, Denmark} 
\affiliation{$^{16}$INFN – Laboratori Nazionali di Frascati, Via E. Fermi 54, P.O. Box 13, 00044 Frascati (RM), Italy}  
\affiliation{$^{17}$CNR - Istituto Struttura della Materia and Elettra-Sincrotrone Trieste, Basovizza Area Science Park, 34149  Trieste, Italy}
\affiliation{$^{18}$RICMASS - Rome International Center for Materials Science - Superstripes, Via dei Sabelli 119A, 00185 Rome, Italy}
\affiliation{$^{19}$INFN, Sezione di Roma, Piazzale Aldo Moro 5, 00185, Rome, Italy}
\affiliation{$^{20}$INGV, Sezione di Bologna, Via Donato Creti 12, 40128 Bologna, Italy} 
\affiliation{$^{21}$INAF - Osservatorio Astronomico di Cagliari, Via della Scienza 5, I-09047, Selargius, Italy} 
\affiliation{$^{22}$Osservatorio Polifunzionale del Chianti, S.P. 101 di Castellina in Chianti, I-50021 Barberino Val d’Elsa, Italy} 
\affiliation{$^{23}$Dipartimento di Fisica e Astronomia, Universit\`a degli Studi di Firenze, Via G. Sansone 1, I-50019 Sesto Fiorentino, Italy}
\affiliation{$^{24}$Precision Mechatronics Laboratory, Aerospace and Mechanical Engineering Department, University of Li\`ege, 9 all\'ee de la d\'ecouverte, 4000 Li\`ege, Belgium} 
\affiliation{$^{25}$BEAMS department, Universit\'e Libre de Bruxelles, 50 av. F.D. Roosevelt, 1050 Brussels, Belgium}
\affiliation{$^{26}$INAF - Osservatorio Astrofisico di Arcetri, Largo E. Fermi, 5, 50121, Florence, Italy} 
\affiliation{$^{27}$INGV, Sezione di Pisa, Via Cesare Battisti 53, 56125, Pisa, Italy} \affiliation{$^{28}$Centre for Cosmology, Particle Physics and Phenomenology (CP3), Universit\'e catholique de Louvain, B-1348 Louvain-la-Neuve, Belgium}
\affiliation{$^{29}$INGV, Sezione di Sismologia e Tettonofisica, Via di Vigna Murata 605, I-00143, Rome, Italy}
\affiliation{$^{30}$International Research School of Planetary Sciences, Viale Pindaro 42, I-65127, Pescara, Italy}
\affiliation{$^{31}$Dipartimento di Ingegneria e Geologia, Universit\`a d’Annunzio, Viale Pindaro 42, I-65127, Pescara, Italy} 
\affiliation{$^{32}$VU University Amsterdam, 1081 HV Amsterdam, Netherlands}
\affiliation{$^{33}$Department of Physics, University of Maryland, College Park, MD 20742, U.S.A.} 
\affiliation{$^{34}$INAF -- Osservatorio di Astrofisica e Scienza dello Spazio di Bologna, Via Piero Gobetti 93/3, I-40129 Bologna, Italy} 
\affiliation{$^{35}$INFN, Sezione di Genova,  Via Dodecaneso 33, I-16146 Genova, Italy} 
\affiliation{$^{36}$Dipartimento di Fisica, Universit\`a di Genova, Via Dodecaneso 33, I-16146 Genova, Italy}
\affiliation{$^{37}$Dipartimento di Fisica e Astronomia  (DIFA), Universit\`a di Bologna, Via Irnerio, 46, I-40126 Bologna, Italy} 

\date{\today}

\begin{abstract}
Monitoring of vibrational eigenmodes of an elastic body excited by gravitational waves was one of the first concepts proposed for the detection of gravitational waves. At laboratory scale, these experiments became known as resonant-bar detectors first developed by Joseph Weber in the 1960s. Due to the dimensions of these bars, the targeted signal frequencies were in the kHz range. Weber also pointed out that monitoring of vibrations of Earth or Moon could reveal gravitational waves in the mHz band. His Lunar Surface Gravimeter experiment deployed on the Moon by the Apollo 17 crew had a technical failure rendering the data useless. In this article, we revisit the idea and propose a Lunar Gravitational-Wave Antenna (LGWA). We find that LGWA could become an important partner observatory for joint observations with the space-borne, laser-interferometric detector LISA, and at the same time contribute an independent science case due to LGWA's unique features. Technical challenges need to be overcome for the deployment of the experiment, and development of inertial vibration sensor technology lays out a future path for this exciting detector concept.
\end{abstract}

\maketitle

\section{Introduction}
\label{sec:intro}
With the first detection of gravitational waves (GWs) from a merging black-hole binary finally achieved in 2015 \cite{AbEA2016a}, the field of gravitational-wave astronomy is still in its infancy. It has already revolutionized multi-messenger astrophysics \cite{AbEA2017d,AbEA2017e}, led to important constraints on stellar-evolution models \cite{AbEA2019c,AbEA2020}, and has put strong constraints on alternative theories of gravity \cite{BaEA2017}, but its main impact on cosmology \cite{AbEA2009,AbEA2017f} and fundamental physics \cite{ArEA2011,YYP2016,AbEA2018,CFK2019,AbEA2019b} is yet to come.

We should expect that the vastness of the yet unfulfilled GW science case will manifest itself in a diversification of the detector concepts to cover the entire frequency band from the slowest possible spacetime oscillations observable in our Hubble volume (of order 10$^{-18}\,$Hz) to the kHz region where we expect the highest-frequency signals produced by astrophysical sources. Detector concepts include, from low-to-high observed frequencies, the CMB B-mode polarization experiments \cite{AdEA2014}, the pulsar timing arrays \cite{HoEA2010}, monitoring of stellar oscillations \cite{SiRo2011,Lop2017}, monitoring of planetary oscillations \cite{Dys1969,Ben1983}, laser-interferometric detectors in space \cite{PhEA2004,SaEA2009,LuEA2016,ASEA2017} and ground-based \cite{HaEA2013,CaEA2020,PMN2020,LSC2015,AcEA2015,Sou2016,AkEA2018}. In addition to the observation band, detectors can also be distinguished by their internal configuration, which has an impact on the information that can be extracted from gravitational-wave signals \cite{BiEA1996,PaEA2016,SaEA2012}, most notably the ability to measure the polarization of a GW and to estimate its propagation direction.

It was Pirani who first studied in great detail the observable effects of GWs \cite{Pir1956} and Weber conceived the first detector concept, i.e., the resonant-bar detector \cite{Web1960}. Weber's efforts were essential to start the field of experimental GW detection, which first developed into a global network of resonant-bar detectors \cite{Ron2006}. It was also understood that Earth itself has an elastic response to GWs. A first analysis leading to an upper limit for GW energy passing through the Earth was obtained in 1961 by Forward et al. \cite{FoEA1961}, and a possible detection was claimed 10 years later by Tuman \cite{Tum1971}. The Lunar Surface Gravimeter experiment, brought on the Moon by the Apollo 17 mission in 1972, had the main scientific target of detecting GWs, but a technical failure rendered the data useless \cite{GiEA1977,BLK1979}. In 2008, the feasibility of a lunar GW detector based on Weber's idea was again discussed in light of new technological developments \cite{PaVe2009}. 

Calculations of cross-sections of elastic bodies to GWs were presented in some of the earliest publications on GW detection, but detailed calculations of the coherent GW response were not known until the end of the 1960s. Dyson calculated the response of a homogeneous, elastic halfspace to GWs \cite{Dys1969}, and the formalism to calculate the response of a laterally homogeneous spherical body was developed by Ben-Menahem \cite{Ben1983}. 

Work published in a recent series of papers has demonstrated that scientifically interesting GW sensitivities can be achieved by monitoring seismic fields \cite{CoHa2014,CoHa2014b,CoHa2014c}. The new analyses were based on state-of-the-art detection pipelines developed by the Virgo and LIGO communities and modified to be applied to a network of seismometers or gravimeters monitoring vibrations of the Earth \cite{CoHa2014,CoHa2014b}, or the Moon \cite{CoHa2014c}. The Dyson half-space response was exploited in \cite{CoHa2014,CoHa2014c}, which is a valid model at higher frequencies where individual normal modes cannot be resolved anymore, and the Ben-Menahem equations for normal-mode excitation were used in \cite{CoHa2014b}. In terms of GW energy density, the new constraints were better by more than 10 orders of magnitude compared to previous limits obtained from high-precision laboratory experiments. The sensitivity limitations of these studies were a product of a trade-off between selecting the quietest data stretches to minimize seismic correlations and using as much data as possible to minimize statistical errors. This trade-off has an optimum when the seismic correlations match the statistical errors. For Moon data, relatively few data stretches had to be excluded from the reported GW analyses, and since the stationary noise is also significantly quieter on the Moon compared to Earth, which reduces the statistical errors, a much better GW sensitivity resulted with Moon data.

With the Lunar Gravitational-Wave Antenna (LGWA), we propose to deploy an array of high-end seismometers on the Moon to monitor normal modes of the Moon in the frequency band 1\,mHz –- 1\,Hz excited by GWs \cite{HaEA2020b}. Several properties of the Moon make it an ideal candidate as GW detector: (1) it is the closest body to Earth, (2) it is large, (3) it lacks an atmosphere and ocean, and (4) it has a much lower seismic activity than Earth. Moonquakes and meteoroid impacts occur (several thousand were identified by the Apollo Lunar Surface Experiments Package (ALSEP) \cite{NaEA1981,LoMo1993}), but the magnitudes of these events are all minor. Most important is that the ambient field, i.e., a stationary background to the Moonquakes and impacts, is so quiet that it was not possible to observe it with ALSEP. 

Using technology with high readiness level, the experiment could run within the next decade, i.e., together with the LISA mission \cite{ASEA2017} or the third-generation, ground-based detectors Einstein Telescope \cite{PuEA2010} and Cosmic Explorer  \cite{ReEA2019}. Since its performance would improve with the development of new vibration sensing technology, it has the potential to become a lasting contribution to the future GW detector network. We also point out that large-scale, laser-interferometric GW detector concepts have been proposed for the Moon \cite{StEA2020,JaLo2020}.

In section \ref{sec:concept}, we describe in detail the important aspects of the detector concept presenting sensitivity requirements and compatibility with the current state of technology. The components of the LGWA concept are brought together in section \ref{sec:sensitivity} to obtain sensitivity predictions for LGWA following several GW search methods. The LGWA science case, which includes the direct analysis of GW signals as well as of the corresponding electromagnetic counterparts detectable by Earth-based and satellite telescopes, is summarized in section \ref{sec:science}. The deployment and operation of the array poses certain practical challenges, which are outlined in section \ref{sec:implementation}. 

\section{Detector Concept}
\label{sec:concept}
In order to assess the quality of a GW detector concept, it is useful to divide the detector into a readout system, and a response body. We understand the response body as an abstract term, which can refer to a laser beam, suspended test masses, clocks, or an elastic body, which can all be affected by a passing GW. At this level, the description of the response can depend on the coordinate system used to describe the experiment. The readout system consists of the coupling dynamics to the response body and the translation of the GW signal into a human-readable form. The quality of a GW detector concept then hinges on three main criteria:
\begin{itemize}
    \item How strongly does the body respond to GWs? 
    \item How sensitive is the readout system to changes in the response body affected by GWs?
    \item What is the level of environmental or intrinsic random excitation of the response body?
\end{itemize}
\paragraph{Response} In the case of LGWA, the response body is the Moon itself. The strength of the response is described by the following relation between GW strain amplitude $h$ and seismometer displacement signal $\xi$:
\begin{equation}
\xi(f) = \frac{1}{2}h(f)\sum\limits_{n=0}^\infty L_n\frac{-f^2}{f_n^2-f^2+\ii f_n^2/Q_n},
\label{eq:response}
\end{equation}
where $\ii$ is the imaginary unit, and $f$ is the signal frequency. The equation neglects any angular dependence of the response. The sum is over the vibrational quadrupole modes of order $n$ of the Moon characterized by a mode frequency $f_n$ and a quality factor $Q_n$. The parameter $L_n$ describes an effective baseline, which can take a complicated form depending on how the amplitude of a mode varies inside the Moon as a function of the distance to its center. All three parameters depend strongly on the internal structure of the Moon \cite{Ben1983}. It should also be noted that displacements are produced in vertical direction (by spheroidal modes) and in the horizontal (by both, spheroidal and toroidal modes). The excitation of toroidal modes by GWs is strongly suppressed, i.e., toroidal modes cannot be excited at all by GWs in a homogeneous body \cite{BiEA1996}. We neglect toroidal modes in the following.

Note that $\xi$ is not (necessarily) the surface displacement induced by GWs, but the difference of surface displacement and direct seismometer test-mass displacement caused by gravitational fluctuations. It was shown in section 2.1.3 of \cite{Har2019}  that this is the natural dynamical variable in models where gravitational strain produces elastic deformations instead of Newtonian gravity potentials. In these coordinates, the effective baseline $L_n$ does not have to be positive, but it is always real-valued \cite{Ben1983}. 

If the signal frequency $f$ coincides with one of the lower-order normal mode frequencies, which are expected to have Q-values of a few 100 at least, then the response can be approximated as
\begin{equation}
\xi(f_n) \approx \frac{\ii}{2}h(f_n)L_nQ_n.
\end{equation}
This result allows us to calculate order-of-magnitude estimates of displacement signals produced by GWs. With a strain amplitude of $10^{-21}$ as can be expected for loud galactic double white dwarfs DWD \cite{KuEA2018,LaEA2019}, a Q-value of 200, which can be expected for the lowest-order quadrupole modes, and an effective baseline of $L_n=0.6R$ \cite{CoFa1996,CoHa2012b}, where $R=1.7\times 10^{6}\,$m is the radius of the Moon,  one obtains displacement signals of order $\xi\sim 10^{-13}\,$m. This response is similar in magnitude to the response in the LISA detector, where it is achieved with a much longer baseline ($L=2.5\times10^9\,$m), but over a broad band without any resonant amplification factor.

\begin{figure}[ht]
\centering
\includegraphics[width=0.95\columnwidth]{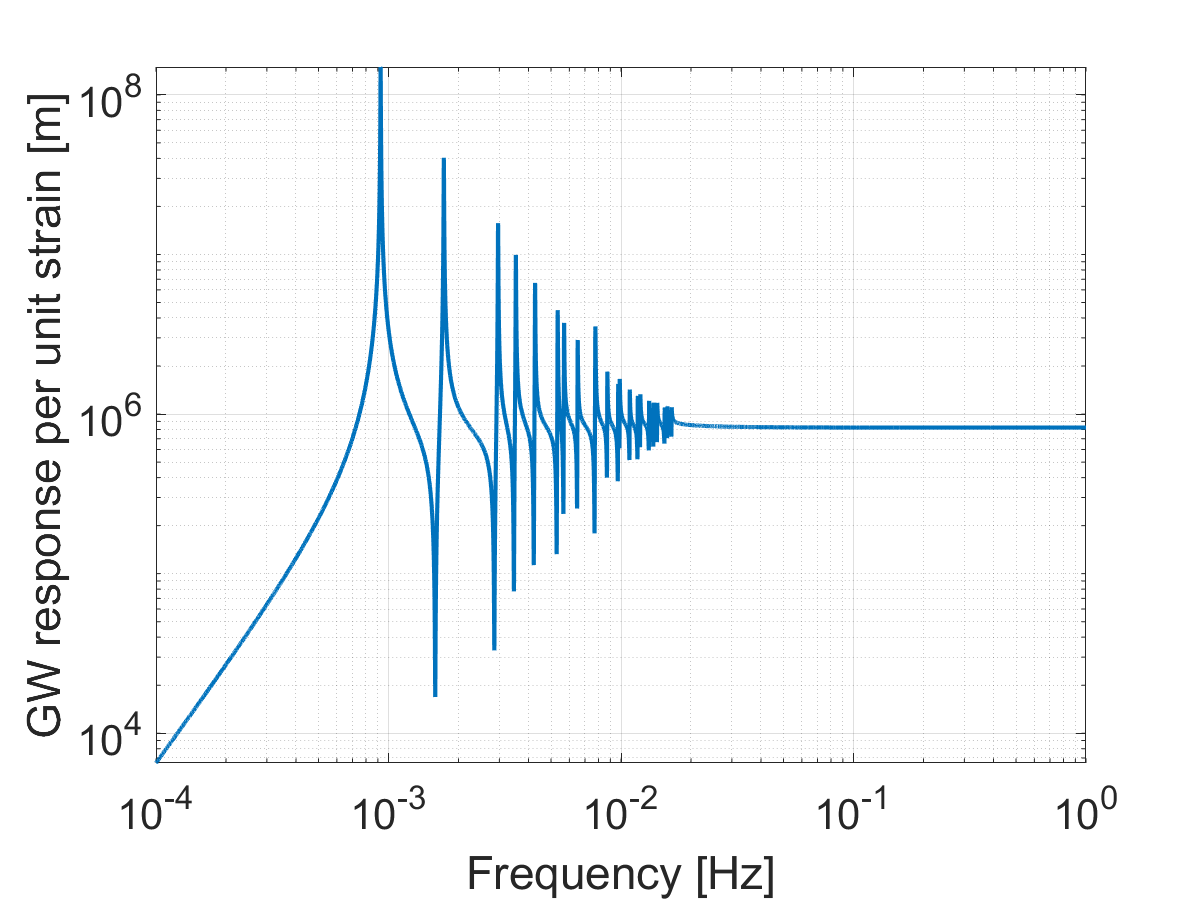}
\caption{Simplified GW response model truncating the normal-mode sum at $n=22$. The model only includes spheroidal quadrupole modes.}
\label{fig:response}
\end{figure}

Another regime of interest is the high-frequency response (around 0.1\,Hz), where it will be difficult to identify excitations of individual normal modes. It turns out that the effective baseline $L_n$ drops very quickly with increasing order $n$ \cite{CoFa1996}. This means that the response can be approximated by truncating the sum in equation (\ref{eq:response}) such that all of the included modes have frequencies well below 0.1\,Hz. The result for a fiducial response model is shown in figure \ref{fig:response}.
The truncated high-frequency response can be approximated as
\begin{equation}
\xi(f\gg f_{\rm max}) = \frac{1}{2}h(f)\sum\limits_{n=0}^{n_{\rm max}} L_n,
\label{eq:respappr}
\end{equation}
where $f_{\rm max}=f_n$ at $n=n_{\rm max}$. The Q-values do not play a role anymore, and the truncated sum over $L_n$ now defines the overall effective baseline. Since in general the baselines $L_n$ can also take negative values \cite{CoHa2014b}, which depends on the Moon's internal structure, there is a partial cancellation of baselines in the sum reducing the high-frequency GW response of the Moon. Intuitively speaking, the question is how the softness of the Moon evolves with increasing frequency. In the limit $f\rightarrow\infty$, the GW response as measured by a seismometer must vanish since the Moon's elastic forces cannot withstand the effective tidal force of a GW. The Dyson response model predicts a $1/f$ approach to this limit \cite{Dys1969}, but details of the geology likely have an important influence on the GW response up to frequencies well above the LGWA band, which means that a simple dependence on frequency should not be expected. The response above 10\,mHz shown in figure \ref{fig:response} is an upper limit, since it assumes that all normal-mode excitations up to order $n=22$ add constructively. The effect of the signs of the baselines $L_n$ on the on-resonance response at lower frequencies is in any case negligible. 

\paragraph{Readout} The second criterion, the quality of the readout system, is defined by the seismometers used to monitor seismic fields. The most sophisticated seismic sensor deployed outside Earth is the SEIS experiment of the Mars Insight mission \cite{LoEA2019}. It achieves an acceleration sensitivity of about $10^{-9}\,\rm (m/s^2)/\sqrt{Hz}$ at mHz frequencies \cite{MiEA2017}. There is significant margin though for sensitivity improvements. The SEIS noise budget contains atmospheric disturbances and noise from thermal processes \cite{MiEA2017}. For this reason, it is expected that the SEIS technology on the Moon would immediately lead to a 10 fold better sensitivity. Another major improvement can be obtained by realizing that seismometers in LGWA do not need to monitor vertical displacement. This means that the stiffness of the mass suspensions can be further reduced \cite{Win2002,BeEA2006} leading to lower suspension-resonance frequencies and improved response of the sensor to ground motion. In addition, substituting the capacitive readout of the test-mass position by an optical readout promises a significant gain in low-frequency sensitivity \cite{BeEA2014,Rderosa2010}. As another solution, one can consider magnetic levitation of a test mass \cite{PaVe2009}. Operating such a system at cryogenic temperatures would be ideal and push the fundamental sensitivity limits of seismic sensing, but since the practical challenges of a cryogenic sensor on the Moon might be too severe, an optical readout of a suspended mass at ambient temperature could be used and achieve superior sensitivities. 
\begin{figure}[ht]
\centering
\includegraphics[width=0.95\columnwidth]{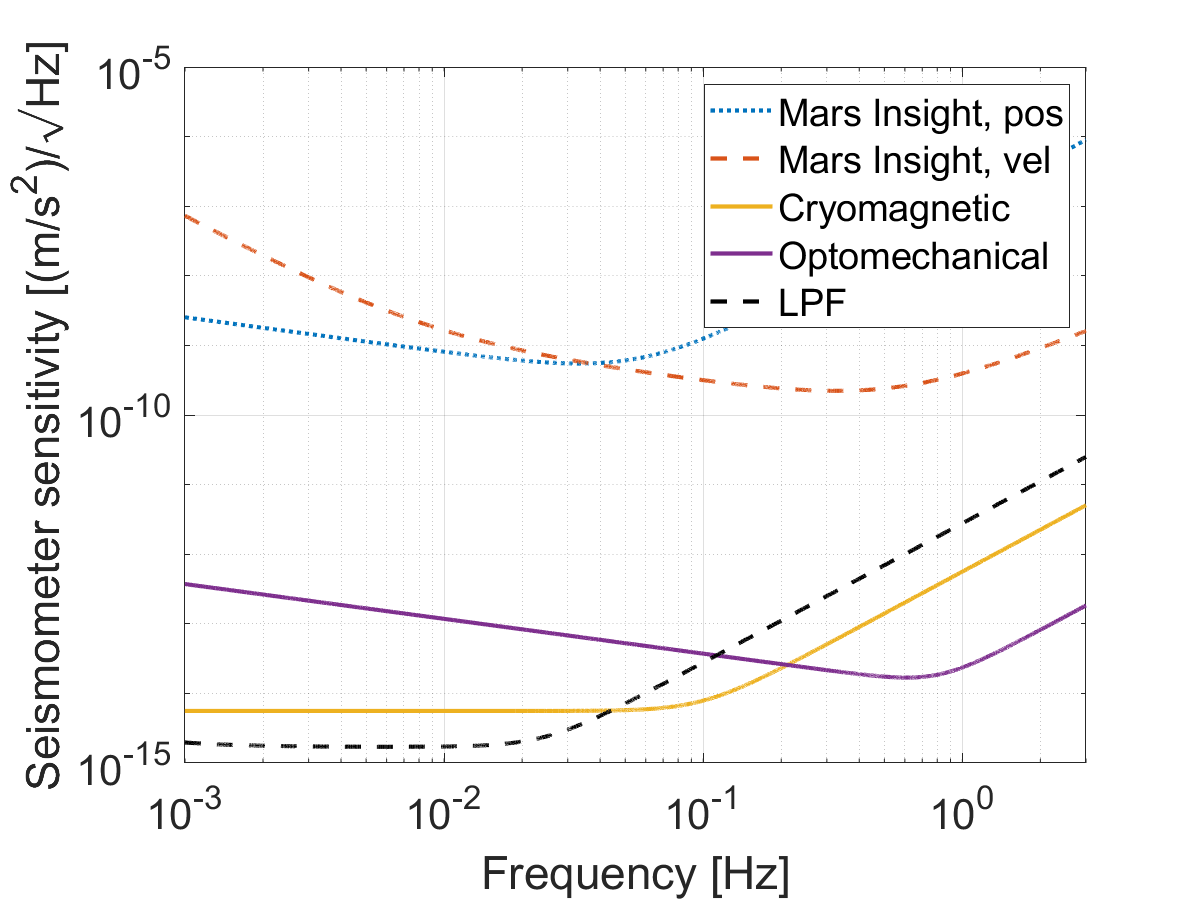}
\caption{LGWA seismometer concepts (cryomagnetic and optomechanical) in comparison with the performance of Mars Insight VBB sensors \cite{MiEA2017}, and with the acceleration sensitivity of the LISA Pathfinder mission \cite{AnEA2018}.}
\label{fig:readout}
\end{figure}
Figure \ref{fig:readout} shows an optomechanical and cryomagnetic seismometer concept for LGWA together with the noise performance of the Mars Insight Very Broadband (VBB) sensors and of the LISA Pathfinder (LPF) mission \cite{ArEA2016,AnEA2018}, which is the best acceleration sensitivity ever achieved at mHz frequencies. The Mars Insight model combines the position and velocity readouts of the suspended masses. Details of the two LGWA concepts can be found in section \ref{sec:sensors}.

\paragraph{Intrinsic noise} The last criterion concerns the intrinsic quietness of the Moon. The small tide signature and the absence of oceans set the background noise level well below the Low Noise Model for the Earth, which was first defined by Peterson in 1993 \cite{Pet1993}. This enhances the capability of seismometers to detect natural events both endogenic (as moonquakes) or exogenic (as gravitational waves). There are four different classes of natural seismic events on the Moon (in addition to artificial impact events):
\begin{itemize}
    \item Deep Moonquakes probably produced by tides;
    \item Shallow Moonquakes a few tens of km below surface;
    \item Thermal quakes;
    \item Meteoroid impact.
\end{itemize}
About 12,500 such events were identified over the course of 9 years and with up to 4 seismometers monitoring in parallel by ALSEP \cite{NaEA1981,KhEA2013}. Even though the number seems high, the annual rate of seismic energy release on the Moon is very small, 4--8 orders of magnitude smaller than on Earth \cite{KhEA2013}. It is expected that the dominant noise background is produced by the impact of meteoroids, which have a relatively high flux for small masses \cite{GHS2011} (only 1700 impacts were strong enough to be identified by ALSEP \cite{KhEA2013}). The value of this background was estimated to be around $5\times10^{-13}\,\rm m/\sqrt{Hz}$ between 0.25\,Hz and 2\,Hz \cite{LoMo1993}, which lies about an order of magnitude above the readout noise of the cryomagnetic concept at these frequencies falling under LGWA instrument noise below 10\,mHz. Our new estimate based on normal-mode response provided in appendix \ref{app:meteoroid} places the meteoroid background below the targeted sensitivity over the entire observation band. Future work must reconcile the two approaches.

\section{LGWA sensitivity}
\label{sec:sensitivity}
\begin{figure}[ht]
\centering
\includegraphics[width=0.95\columnwidth]{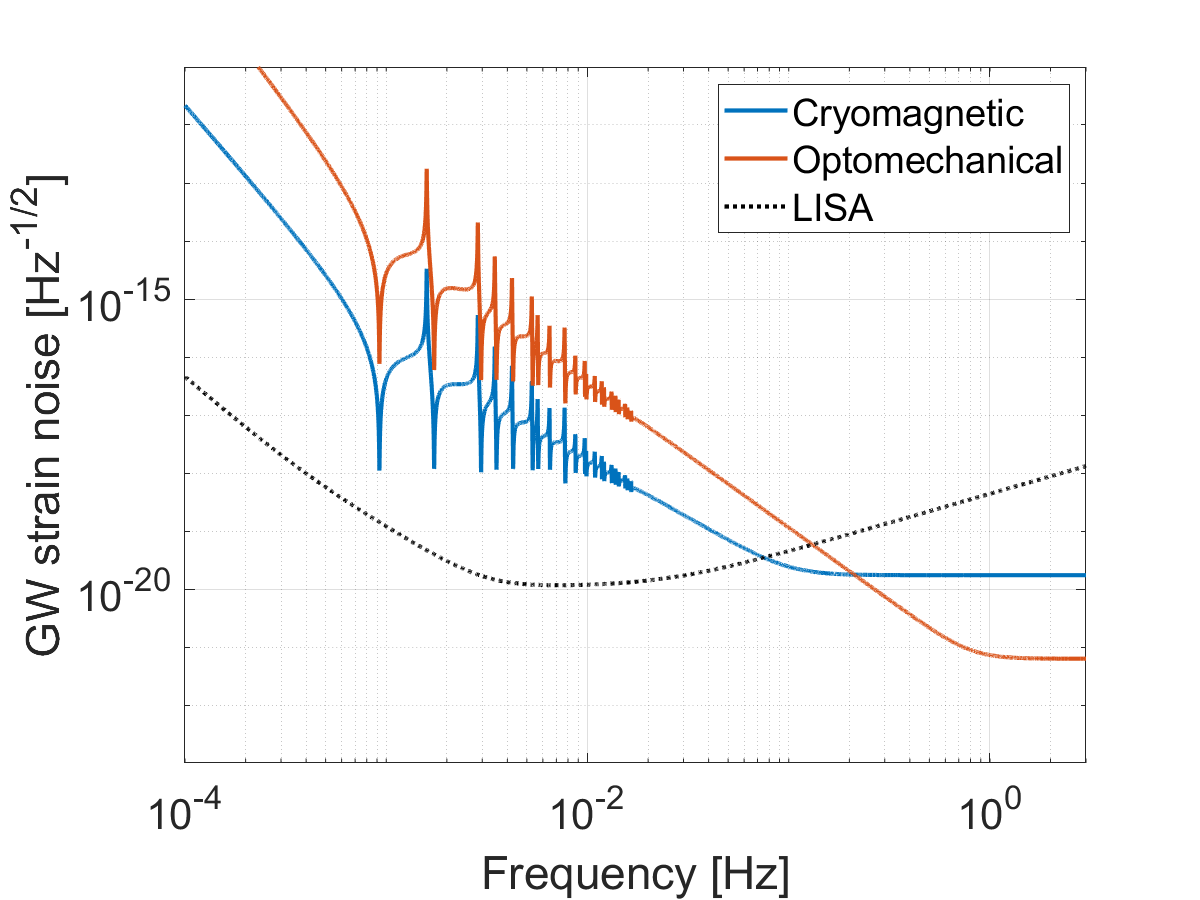}
\caption{Predicted LGWA noise spectral density.}
\label{fig:strainnoise}
\end{figure}

The basic noise spectral density $S_h(f)$ of LGWA in units of GW strain $h$, is obtained dividing the readout noise shown in figure \ref{fig:readout} by the GW response model in equation (\ref{eq:response}). Here, we assume a Q-value of 300 for all quadrupole modes and use a simplified (homogeneous) model of the Moon. The square-root of $S_h(f)$ is shown in figure \ref{fig:strainnoise} together with the noise target of the LISA detector \cite{ASEA2017}.
Interestingly, LGWA has the potential to beat LISA sensitivity above 0.1\,Hz, which might be interesting for observations of DWD mergers. Its noise consists of a serious of narrow-band features inherited from the normal-mode response. However, above 10\,mHz, the response becomes broadband dominated by the above-resonance response of the low-order quadrupole normal modes.

In the initial proposal of LGWA \cite{HaEA2020b}, a division of the mission into phase 1 and phase 2 was foreseen (see section \ref{sec:implementation}). Phase 1 included the deployment of an array of nearby seismometers, while phase 2 referred to the deployment of an additional seismometer on the opposite side of the Moon. Phase 2 is required for GW detections relying on correlation between seismometers as necessary for stochastic GW backgrounds \cite{CoHa2014}. The main issue here is the background noise produced by the meteoroid impacts. Correlating data between two nearby instruments, one can reject contributions from readout noise, but one will likely observe a partial correlation of the meteoroid background noise, which would pose a strong sensitivity limitation to stochastic GW searches. Instead, a near antipodal location of the phase-2 sensor would have very small or negligible correlations from seismic sources, but the correlation of GW signals is still maximal.

A sensitivity with respect to the stochastic GW signal is typically expressed in units of an energy density relative to the critical energy density of the standard cosmological model \cite{All1996}. For an optimally positioned (antipodal) pair of seismometers, one obtains the following sensitivity to measurements of the fractional energy density of GWs:
\begin{equation}
    \Omega(f)=\frac{10\pi^2f^3}{3H_0^2}\sqrt{\frac{1}{fT_{\rm obs}}}S_h(f),
\end{equation}
where $H_0$ is the Hubble constant, and $T_{\rm obs}$ is the total observation time. The factor $fT_{\rm obs}$ represents the number of averages one can maximally do to estimate the cross-spectral density at frequency $f$ between the two seismometers \cite{ShHa2020}.
\begin{figure}[ht]
\centering
\includegraphics[width=0.95\columnwidth]{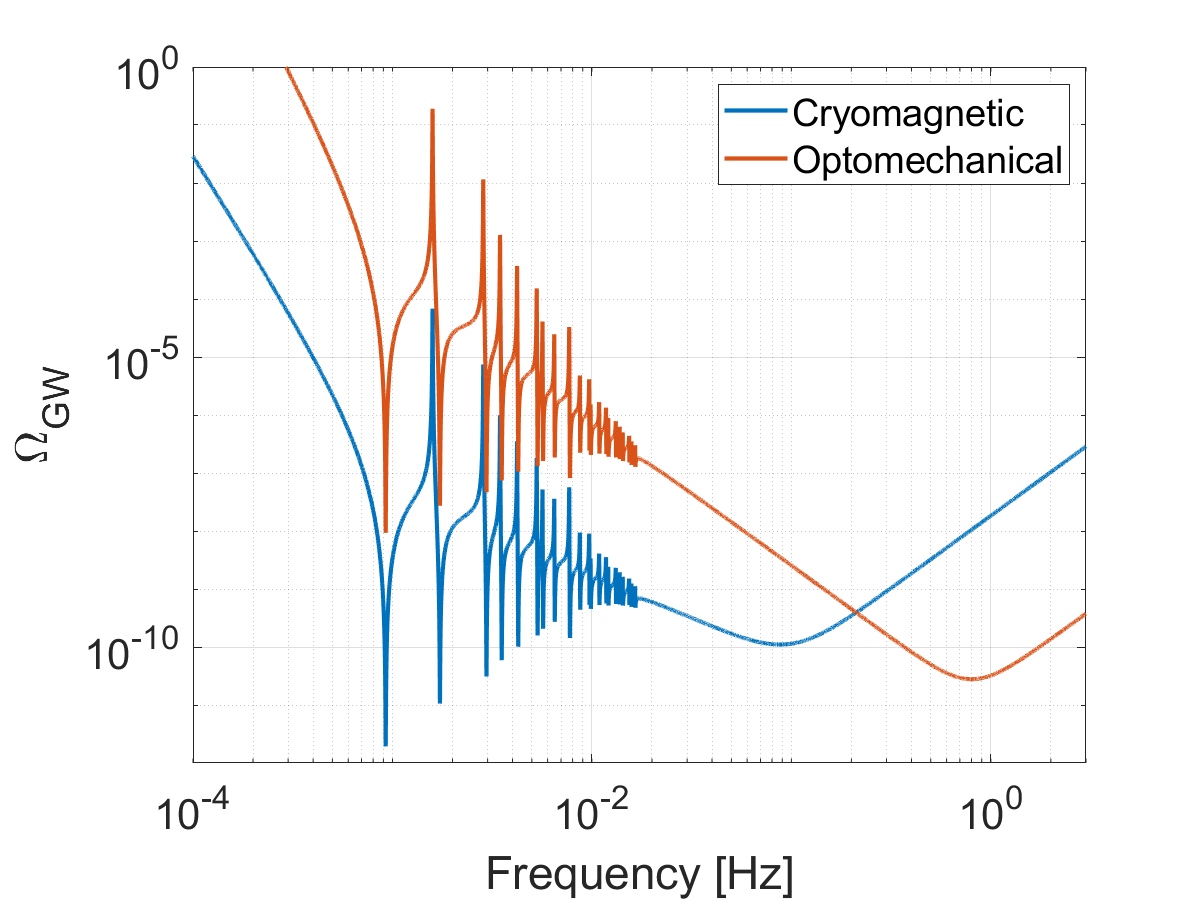}
\caption{Predicted LGWA sensitivity to stochastic GW backgrounds.}
\label{fig:stochastic}
\end{figure}
The resulting noise prediction for measurements of GW energy density is shown in figure \ref{fig:stochastic} with a total observation time of 3 years.

\begin{figure}[ht]
\centering
\includegraphics[width=0.95\columnwidth]{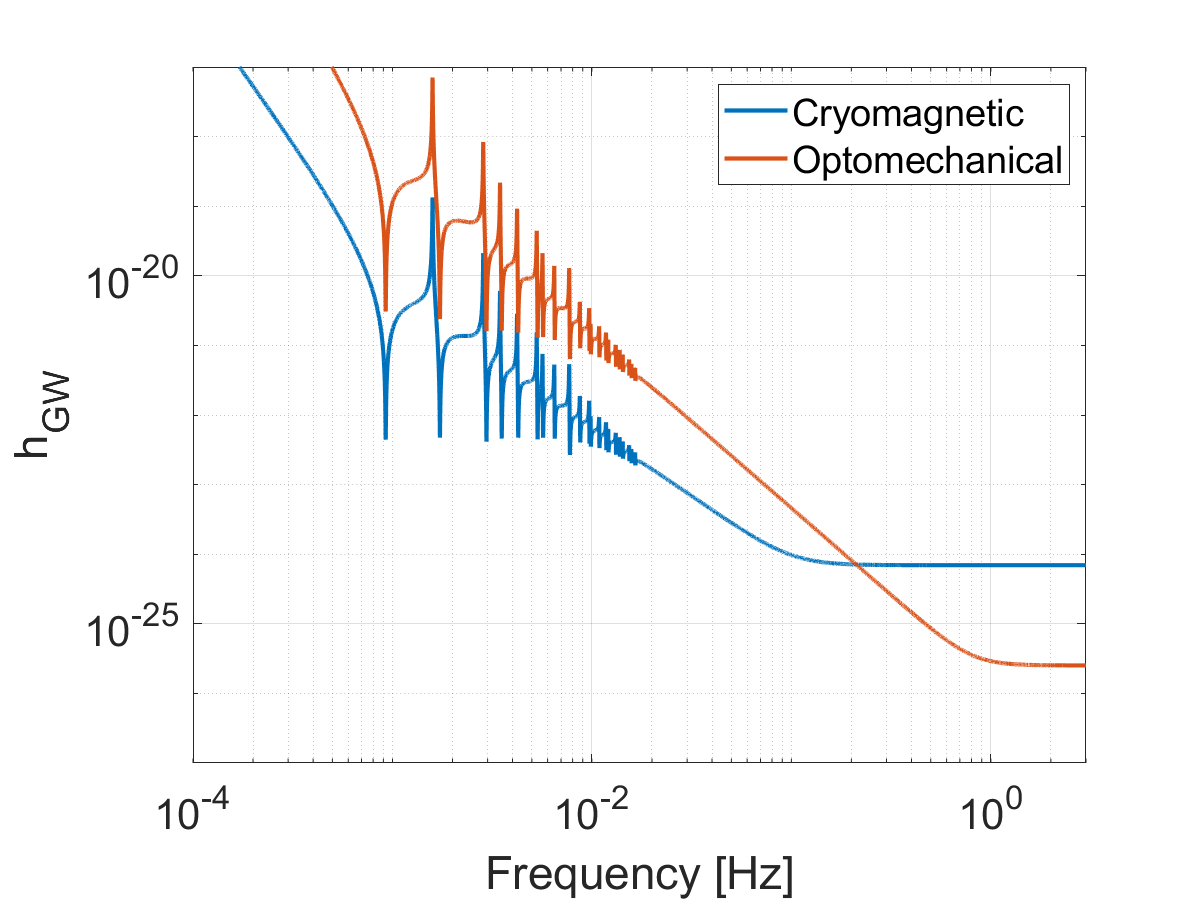}
\caption{Predicted LGWA sensitivity to near-monotonic GW signals.}
\label{fig:amplitude}
\end{figure}

The most sensitive searches of GWs could be carried out by matching the (still unknown) normal-mode frequencies of the Moon with catalogues of double white dwarfs (DWDs) \cite{KuEA2018}. These binaries emit GWs at known and slowly increasing frequencies and their waveforms can be predicted. Alternatively, one could search over entire regions of the DWD parameter space to detect unknown DWD systems, which is the standard matched-filter search carried out for the current ground-based GW detectors. In either case, the minimal value of the GW amplitude that can be resolved is \cite{HaEA2013}
\begin{equation}
    h_0=(S_h(f)/(2T_{\rm obs}))^{1/2}\cdot \rm SNR_0,
    \label{eq:strainT}
\end{equation}
where SNR$_0$ is the signal-to-noise threshold for a GW detection. The noise level of this measurement, i.e., the last equation with SNR$_0=1$, is shown in figure \ref{fig:amplitude} for a 5\,yr observation time. 
In order to be able to interpret this result as GW-amplitude noise, one must assume that the frequency of the signal does not change significantly over periods of a few years as is the case for all mHz DWDs. Merger times of massive black-hole binaries would be shorter, which means that within a year, such a source could ring up normal modes in a temporal sequence from low to high frequencies. The analysis of LGWA data above 10\,mHz can follow conventional methods with respect to all types of GW signals since the sensitivity loses its peaked features.

\section{Implementation and technologies}
\label{sec:implementation}

\subsection{Lunar surface environment}
Several environmental factors can be important for LGWA including lunar dust \cite{GHS2011}, lunar surface temperature \cite{WiEA2017}, cosmic rays and charges \cite{StEA2006,JoEA2014}, and the surface magnetic field \cite{LiEA1998}. 

The dust cloud surrounding the Moon can be important for the operation of sensitive equipment. Overheating of the Apollo 11 seismometer was attributed to dust deposition \cite{OBr2012} (likely connected to the launch of the lunar module). It is certainly important to keep in mind for the instrument design that sensitive parts are not exposed to the environment. 

Surface temperatures during lunar nights can fall close to 100\,K and rise up to 400\,K during days \cite{WiEA2017}. As will be explained in section \ref{sec:sensors}, of particular interest is the south pole of the Moon with its permanently shadowed regions (PSRs) where temperatures well below 100\,K can be found \cite{PaEA2010}. These could be used as natural cryostats for a lunar seismometer. 
\begin{figure}[ht]
\centering
\includegraphics[width=0.95\columnwidth]{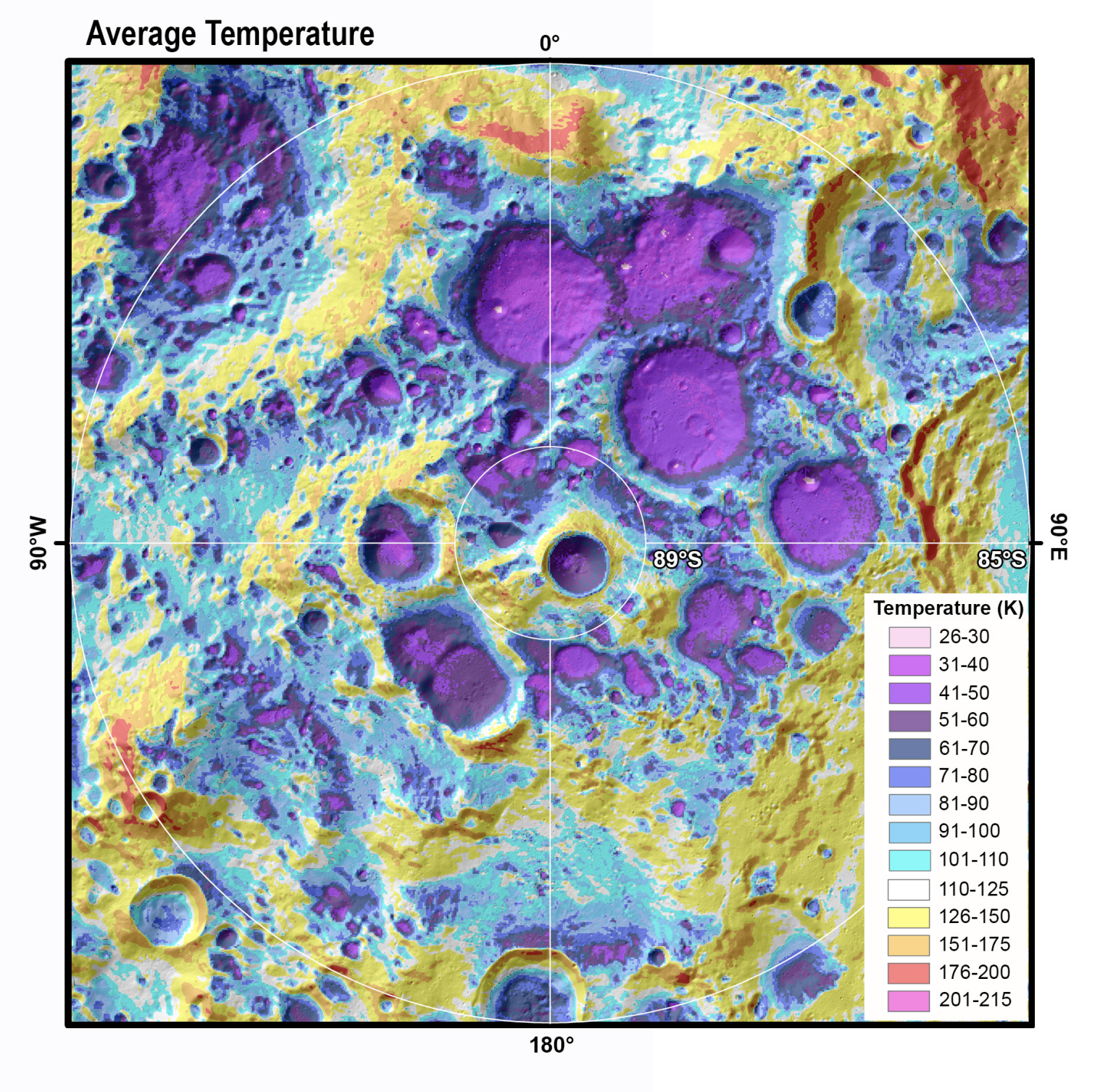}
\caption{Average surface temperatures at the lunar south pole \cite{PaEA2010,Sto2019}.}
\label{fig:temperatures}
\end{figure}

Radiation, for example in the form of galactic cosmic rays, can damage electronics \cite{Duz2005}. Radiation hardening is a common technique to make electronics more resistant against ionizing radiation. In addition, cosmic rays can lead to continuous charging of the lunar regolith, which can develop significant electric field strengths \cite{JoEA2014}. 

Finally, seismic sensors as required for LGWA can be susceptible to fluctuations of the magnetic field \cite{Ack2015}. Such fluctuations have not been characterized in detail on the lunar surface yet, but past measurements indicate that the magnetic field is about a factor 1000 weaker than on Earth \cite{LiEA1998}. Even if this factor extended to the magnetic fluctuations, due to the extreme sensitivity of the LGWA seismometers to external forces, it would be necessary to design the instrument considering magnetic couplings.

\subsection{Deployment}
The concept foresees the deployment of several seismic stations on the Moon operating for several years. The goal is the observation of GWs between about 1\,mHz and 1\,Hz. In the original LGWA proposal \cite{HaEA2020b}, a phase-1 deployment of 4 seismometers to form a km-scale array near the north-western edge of Oceanus Procellarum is followed by a phase-2 deployment of a seismometer on the back side of the Moon at an antipodal location with respect to the phase-1 array. The phase-1 site was chosen to be far from locations of potential interest for other missions to avoid problems with excess seismic noise produced by these activities. The phase-2 site was chosen to have minimal seismic correlations between phase-1 and phase-2 seismometers, but at the same time, the antipodal location means that correlations due to GWs would be maximal \cite{CoHa2014}. 

In fact, the most sensitive past studies were based on correlation measurements between seismometers \cite{CoHa2014,CoHa2014b,CoHa2014c}. Such measurements are ideal to reveal astrophysical or primordial stochastic GW backgrounds. Seismic correlations were either minimized by choosing antipodal seismometer pairs \cite{CoHa2014} or even by correlating data between one seismometer on the Moon and another on Earth \cite{CoHa2014c}. However, such constellations would be less useful for other types of GW searches not relying on correlation measurements between seismometers, e.g., searches for modeled signals or for unmodeled GW transients. Instead, the km-scale, phase-1 array configuration offers a great advantage in these cases compared to a collection of widely separated seismometers as explained in the following.

Assuming that the continuous seismic background oscillations on the Moon are too weak to be observed by LGWA (see appendix \ref{app:meteoroid} for an estimate of the background from meteoroid impact), the sensitivity of LGWA is generally limited by seismometer self-noise. However, there can still be individual, transient seismic events dominating the signal occasionally. At LGWA's sensitivity level, the rate of such transients is unknown, e.g., the Apollo seismometers were less sensitive and could only see events with larger magnitude compared to what LGWA would be able to see. There could well be thousands of significant seismic events per year. Now, it would be sufficient to be able to identify seismic events and subtract them from the data to restore LGWA's full sensitivity potential. A similar (but more complicated) technique is currently being developed for the ground-based GW detectors to enhance their sensitivity, called Newtonian-noise cancellation \cite{CoEA2018a,TrEA2019,Har2019,HaEA2020}. It requires an array of closely spaced seismometers to perform a coherent subtraction of seismic disturbances from a target channel. Therefore, as preventive measure against frequent seismic transients, we propose to deploy a km-scale array of at least 4 seismometers. The optimal diameter of the array used for this purpose still needs to be calculated, but it must be small enough so that data from 3 seismometers can be used to accurately infer the signal at the fourth.  An open question is up to which surface amplitude events need to be subtracted from the data to provide a sufficient amount of transient-free GW data. Subtracting all events, up to the strongest ones, would require a larger dynamic range --- or more precisely, a larger linear response range --- of the seismometers, which would be an additional challenge for the instrument development.

\begin{figure}[ht]
\centering
\includegraphics[width=0.95\columnwidth]{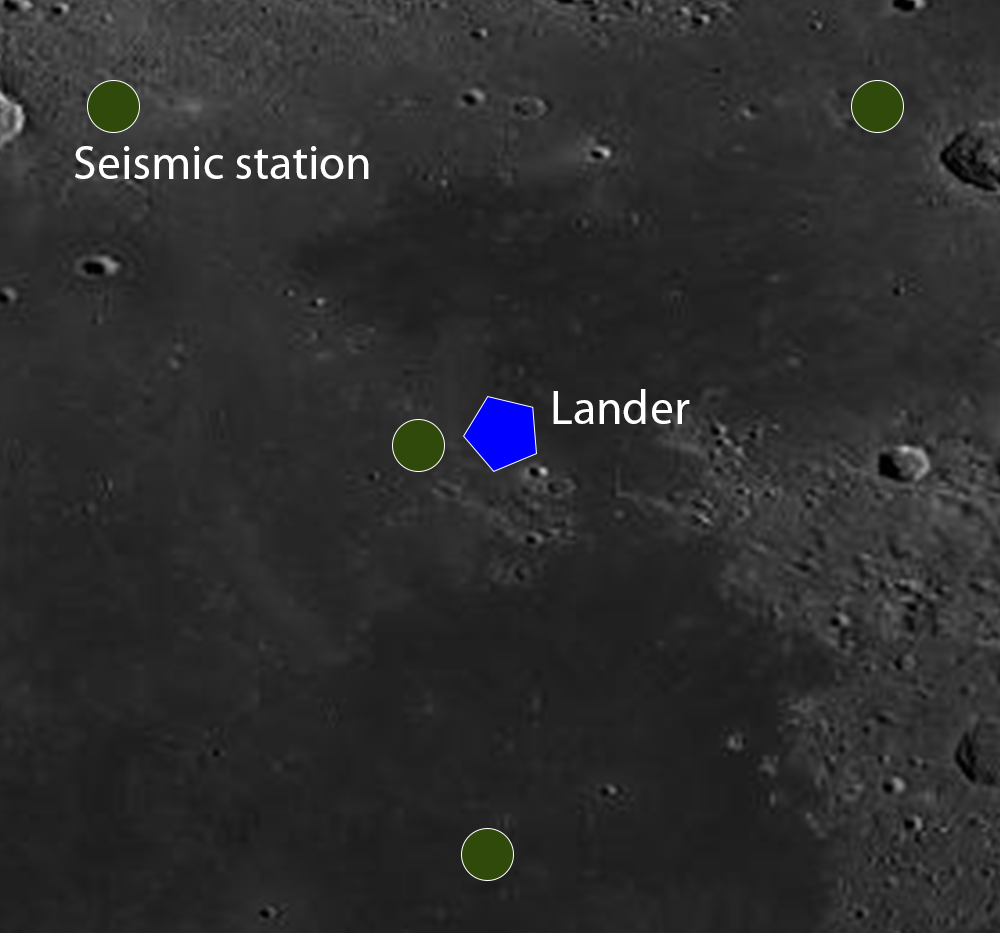}
\caption{Configuration of the km-scale LGWA array. Lander position is indicated for the case when all seismic stations are deployed from a single landing site. Separate landings are possible.}
\label{fig:array}
\end{figure}
An important question is how the seismic stations are to be deployed. Since researchers in the US have already developed a robotic lander concept (Lunette) for environmental monitoring stations \cite{ElAl2010}, a possible solution is to deploy the LGWA stations with individual robotic landers. The challenge here is to achieve the required landing accuracy. For a km-scale array, landing accuracy should be better than 100\,m so that the configuration is not dominated by landing deviations. A few 100\,m landing accuracy or better is feasible with a single landing \cite{SiEA2014}. However, since landing accuracy is only required for the relative positions of seismic stations, we propose to have a first station deployed together with a beacon, which can be used in subsequent landings for much improved landing accuracy relative to the beacon position (well below 100\,m).  

Alternatively, deployment from a single landing site using drones or rovers can be considered. The use of drones would have the advantage that deployment can be very quick and less dependent on the terrain. However, a novel propulsion and steering system would be required to operate in vacuum and to transport a load of about 20\,kg (which is our current estimate of the mass of a seismic station). Rover technology is more advanced, but deployment of several seismic stations would be slow and there is an increased risk that some target locations cannot be reached easily. A navigation system would have to be realized. It could be based on SLAM (simultaneous localization and mapping) techniques \cite{SLL2005,GuEA2018} using data from a camera feed. This would benefit from extremely wide-field optical stereoscopic cameras as investigated by some of us \cite{PeEA2020}.

In all cases, relative positions of seismic stations need to be measured after deployment with errors smaller than 10\,m to provide the required accuracy of travel-time estimates of seismic waves.

\subsection{Lunar model and data calibration}
 Essential for the analysis of LGWA data is an accurate model of the lunar interior in terms of its elastic properties including shear and compressional seismic speeds, and mass density. These models are typically represented as a function of radius approximating Earth or Moon as a spherical, laterally homogeneous body. They have been employed successfully to explain the observed frequencies and Q-values of Earth's normal modes and the arrival times of seismic phases (or the other way round, the models were first inferred from these observations) \cite{MoKe1996}. A ``very'' preliminary lunar model was calculated by Garcia et al \cite{GaEA2011}. Such models are required to simulate normal modes, which form the basis of a calibration of displacement signals observed with LGWA to infer the corresponding GW signals. The better our understanding of Moon's interior, the better will be the calibration, and the lower will be the impact of calibration errors on GW parameter estimation. 

For the LGWA concept to work, the continuous hum of lunar normal modes --- as has been observed for numerous modes on Earth \cite{SNF1998} --- needs to be weak enough. We estimate in appendix \ref{app:meteoroid} that the continuous hum produced by meteoroid impacts lies about 3 -- 4 orders of magnitude below the targeted noise level of the seismic instruments shown in figure \ref{fig:readout}. This would imply that potentially observable normal-mode excitations, especially in the mHz band, are expected to be rare, transient events described by an exponential ring-down following the initial excitation. These transients would add to an extremely weak continuous hum. Potential observations of these transients together with observations of seismic phases from lunar seismic events by a lunar seismic array can be used to infer a model of Moon's interior. 

Array configurations suitable for these measurements are not necessarily ideal for GW observations. In order to infer Moon's internal structure, seismic stations should be widely separated as planned for the Lunar Geophysical Network \cite{WeEA2020}, while we motivated in the previous section that LGWA should ideally be composed of km-scale arrays to be able to distinguish detectable seismic events from instrument noise and GW signals, and to subtract them from the seismic data. Important improvements of our understanding of Moon's interior can already be made based on data collected with VBB seismometers similar in sensitivity to the Mars Insight seismic sensors. Therefore, these efforts can start right away with instrumentation of advanced technological readiness as part of lunar geophysical missions. 

\subsection{Seismic sensors}
\label{sec:sensors}
In the following, we confront two seismometer concepts: (1) the optomechanical concept based on the mechanical suspension of a test mass and optical readout, and (2) the cryomagnetic concept based on a magnetic levitation of the test mass, and SQUID (superconducting quantum-interference device) readout. The optomechanical concept can in principle operate at any temperature, but cold temperatures are preferred for mechanical stability and reduced thermal noise. The baseline design of the cryomagnetic concept requires a temperature of 9\,K, but a possible realization with high-temperature superconductors should be investigated.

The sensors are meant to monitor horizontal displacement, where soft suspensions with low resonance frequencies can be realized more easily. We require for both sensor concepts that the suspension resonance frequency is $\omega_{\rm s}/(2\pi)=0.01\,$Hz. In compact systems as required for a deployment on the Moon, this can be achieved in mechanical systems making use of spring-antispring dynamics \cite{Win2002}, while compactness does not have a direct impact on the dynamics of a magnetic levitation. 

We start with the optomechanical concept. Mechanical suspensions in the form of springs, cantilevers, and pendula are the basis of almost all commercial seismometers today. However, optical readout has not yet entered commercial systems widely. Instead, readouts with capacitors and coils are favored. These are easier to implement and provide sufficient sensitivity to terrestrial seismic observations. Nevertheless, optical readout has already been prototyped in commercial broadband seismometers substituting the capacitive readout \cite{ZuEA2010,BeEA2014}. New low-frequency seismometer concepts with the goal to eventually provide superior sensitivities are based on optical readouts \cite{AcEA2006,HeEA2018,MoMa2019,Hei2020}. We believe that the optical readout is the better choice whenever the goal is to push sensitivity limits, and we adopt it as our baseline solution for an optomechanical seismometer concept for LGWA.

The square-root of power spectral density of readout noise, which here means quantum shot noise from light detection in interferometric devices, referred to ground displacement is given by
\begin{equation}
\delta x(\omega)=\frac{\lambda_0}{2\pi}\sqrt{\frac{hc}{\lambda_0P}}\frac{\sqrt{(\omega^2-\omega_{\rm s}^2)^2+(\omega_{\rm s}^2/Q)^2}}{\omega^2},
\label{eq:readout}
\end{equation}
which is about $1.4\times 10^{-14}\,\rm m/\sqrt{Hz}$ for $\omega\gg\omega_{\rm s}$ with $P=20\,$mW laser power at $\lambda_0=1064\,$nm. The value of the quality factor has no significant impact on the readout noise above resonance frequency.

The suspension-thermal noise with respect to measurements of ground acceleration ($\delta a(\omega)=\omega^2\delta x(\omega)$) takes the form \cite{HaML2017}
\begin{equation}
\delta a(\omega)=\sqrt{\frac{4k_{\rm B}T}{mQ}\frac{\omega_{\rm s}^2}{\omega}}.
\label{eq:thermal}
\end{equation}
Its value is $4\times 10^{-13}\,\rm (m/s^2)/\sqrt{Hz}$ at 1\,mHz assuming a suspended mass of $m=1\,$kg, a temperature of $T=40$\,K as can be found in PSRs, and a quality factor of $Q=10^4$. It is important to point out that the ratio $\omega_{\rm s}^2/Q$ is largely independent of the resonance frequency. It is mostly determined by material properties and mechanical stress in the bending parts of the suspension \cite{HaML2017}. In other words, achieving a low resonance frequency is not primarily important for the thermal noise, but above all to increase the response of the seismometer to ground motion, which helps to reduce readout noise as given in equation (\ref{eq:readout}). 

Next, we discuss the cryomagnetic concept. As cryomagnetic concept, we understand a magnetic levitation system implementing superconductors in the readout system as well as for the levitation coil and test mass. The readout is done using a SQUID. Such systems already form part of the most sensitive gravity-gradient sensors \cite{MPC2002,GrEA2017}, and similar, but simpler systems can for example be found in superconducting gravimeters of the Global Geodynamics Project \cite{Goo1999,CrHi2010}. It is expected that relatively high Q-values of order $10^6$ or potentially even higher can be achieved even with low resonance frequencies, which gives the cryomagnetic concept an advantage concerning thermal noise. We assume a value of $Q=10^6$ in the following.

Adding SQUID readout noise and thermal noise, the power spectral density of the fundamental instrumental noise of the cryomagnetic concept is given by \cite{PaVe2009}
\begin{equation}
\begin{split}
S_x(\omega)&=\frac{2E_A}{m\omega_{\rm s}^2\beta\eta}\frac{(\omega^2-\omega_{\rm s}^2)^2+(\omega_{\rm s}\omega/Q)^2}{\omega^4} \\
&\quad+\frac{1}{\omega^4}\frac{4k_{\rm B}T}{mQ}\omega_{\rm s}.
\end{split}
\end{equation}
Here, $E_A=1000\hbar$ is the SQUID energy resolution, and $\beta$ and $\eta$ are the electromechanical energy coupling and energy coupling efficiency from circuit to SQUID with $\beta\eta=0.25$. More recently, a modification of the SQUID readout was suggested by implementing an LC-bridge transducer, which could potentially overcome limitations of the readout noise \cite{PaEA2016}. Mostly to avoid the introduction of high-temperature superconductors in our cryomagnetic concept, which are not well explored for levitation systems, we assume that the system is kept cooled at 9\,K. This, of course, can be considered a major hurdle for applications on the Moon, which is why exploration of high-temperature (e.g., 40\,K) solutions is necessary.

Comparing the two concepts, the optomechanical system has an advantage at higher frequencies since relatively low laser power can already achieve extremely low readout noise. It is difficult to imagine that SQUID readout noise can be lowered much relative to the spectrum given in figure \ref{fig:readout}. The cryomagnetic concept is predicted to have a better performance in the thermal-noise limited band. This is not only because of the more optimistic assumptions about achievable Q-factors, but also because of the fact that energy dissipation in magnetic systems is expected to be dominated by viscous damping. 

Generally, it would be possible to combine features of the two concepts to create new concepts. For example, a magnetic levitation could be combined with optical readout. However, we think that the two chosen concepts are the most consistent ones since they either fully exploit the benefit of cryo-temperatures, or are fully compatible with any ambient temperature. It is also conceivable that the cryomagnetic concept would work at ambient lunar temperatures, as long as it is deployed in one of the PSRs, where temperatures well below 100\,K can be found, and then implementing high-temperature (type II) superconductors.

Needless to say, other instrumental noise exists in these concepts, as for example laser-frequency noise, other forms of thermal noise, electronics noise, or coupling to fluctuations of ambient electromagnetic fields, but we believe that these can be suppressed more easily by design and precise engineering. Realizing the targeted Q-value and suspension resonance frequency of any of the two concepts is the main challenge. 

\subsection{Communication and positioning}
There is an enormous effort to create important communication and positioning infrastructure for future missions to the Moon. One idea that has been discussed since long is the usage of the terrestrial GNSS for lunar missions \cite{DeEA2020}. Another interesting opportunity has emerged with the introduction of microsatellites and CubeSats for inter-planetary missions. A CubeSat constellation was proposed to form a Lunar Global Positioning System \cite{BaEA2012a}. While the first phase of the LGWA experiment would not require such infrastructure, subsequent phases especially with station deployments on the backside of the Moon would depend on it for accurate navigation. 

In general, deployment at sites without continuous Earth visibility would require a novel data link based on satellites in lunar orbit. In order to estimate the requirements of data-transmission systems, we provide an estimate of the total instrument data rate and volume for a phase-1 LGWA array with 4 stations in table \ref{tab:data}. 
\begin{table}[ht!]
\begin{tabular}{|p{4cm}|p{1.8cm}|l|l|}
\hline
Parameter &	Multiplier & Data &	Unit \\
\hline
Signal BW to be sampled & &	10 & Hz \\
\hline
Sampling BW (Nyquist) & & 20 & Hz \\
\hline
Bits per sample & & 16 & \\
\hline
Data rate per seismic sensor & & 320 & bits/s \\
\hline
Data volume per day per seismic station & 2 $\times$ 86400\,s &	55,296 & kbits \\
\hline
Total data volume per day & 4 stations & 221,184 & kbits \\
\hline
\end{tabular}
\caption{Calculation of data-transmission rates for LGWA. BW: bandwidth.}
\label{tab:data}
\end{table}

Here, we assume to have 2 horizontal seismic channels per seismic station, and 4 seismic stations in total. Each seismic channel is sampled at 20\,Hz. Based on these numbers, the total data-transmission rate would be about 250\,Mbits per day. Such transmission rates can in principle be achieved easily today even with direct transmission of sensors to the Earth if power is available. The Lunar Atmosphere and Dust Environment Explorer (LADEE) in lunar orbit contained the Lunar Laser Communication Demonstration (LLCD), which demonstrated download transmission rates of about 600\,Mbit/s with a total power consumption of $\sim90\,$W \cite{BoRo2014}. More conventional X-band communication can achieve a few Mbit/s between Earth and Moon \cite{ZhEA2019}, while a $\sim 4\,$W X-band transmitter on CubeSats could achieve several 10\,kbit/s \cite{ScEA2017}, which would still be enough to transmit the total data volume of LGWA with 4 stations. Future private Moon satellite constellations under development by Commstar Space Communications in collaboration with Thales Alenia Space, or the Lunar Pathfinder by SSTL (Surrey Satellite Technology Limited) will provide both communication and navigation services, which can be used by LGWA to facilitate its development.

\section{LGWA Science}
\label{sec:science}

\subsection{Aspects of data analyses}
\label{sec:aspects}
LGWA has a few unique features compared to other detectors, which requires specialized data-analysis techniques, and which give unique capabilities to this concept:
\begin{itemize}
    \item LGWA, due to Moon's rotation with a period of 27.3 days, has a unique temporal evolution of its antenna pattern among all GW detector concepts;
    \item Over time, LGWA can be extended to a distributed array over the Moon's surface, allowing for precise GW polarization measurements \cite{WaPa1976,BiEA1996};
    \item As shown in figure \ref{fig:response}, LGWA has the characteristics of a resonant-bar antenna at frequencies up to 10\,mHz and transitions into a broadband detector above 10\,mHz; 
    \item As shown in figure \ref{fig:strainnoise}, LGWA has the potential to become the most sensitive GW detector in the band 0.1\,Hz to 1\,Hz until new detector concepts like DECIGO \cite{SaEA2017} or BBO \cite{PhEA2004} start operation. 
\end{itemize}
The first point is relevant to the estimation of certain GW signal parameters, especially sky location, provided that the GW signal lasts for long enough as would be the case, for example, with less massive compact binaries including DWD or binary neutron stars (BNS) at mHz frequencies, and GWs from spinning neutron stars. In these cases, the observed signal amplitude experiences a temporal modulation, which can be exploited for parameter estimation \cite{WeCh2010,GrHa2020}. 

The ability of an extended array of seismic sensors to measure polarizations of GWs would benefit parameter estimation. For example, in the case of compact binaries, the degeneracy between extrinsic source parameters like orbital inclination, polarization angle, sky location and distance can be broken, further assisted by temporal variations in antenna patterns as explained earlier. Furthermore, antipodal pairs of seismometers are the ideal configuration for stochastic GW searches since seismic correlations are expected to be small. In this configuration, the two seismometers have the largest possible distance to each other, and at the same time, the overlap-reduction function that describes correlations of GW signals as a function of relative position and orientation is maximal \cite{CoHa2014,CoHa2014c}.

The third point determines what data-analysis techniques need to be used. For example, a massive binary black hole (BBH) would ring up normal-mode resonances from low to high frequencies with a well defined beat (beat in the musical sense). For such signals, below 10\,mHz, information can only be extracted from the beat and the loudness of each ring. Less massive binaries below 10\,mHz, i.e., binaries with negligible frequency evolution, need to hit the right frequencies, i.e., the normal-mode resonances, and will most likely remain undetected off-resonance. Above 10\,mHz, standard data-analysis techniques can be applied as known for all broadband GW detectors. 

The fourth point in the list emphasizes an important complementarity of LGWA's role in a multiband detector network, which will be further elaborated in section \ref{sec:complement}.

\subsection{Binaries}
\subsubsection{Signal modeling}
For the analyses in the following sections, GW amplitudes need to be estimated for various observation and modeling scenarios. So, we start with the characterization of a binary inspiral. Solar-mass binaries in the LGWA band move slowly in frequency. This can be characterized by the time it takes the binary to merge \cite{HaEA2013}
\begin{equation}
T_{\rm insp}=1.4\times 10^3\,{\rm yr}\left(\frac{\mathcal M_{\rm c}}{2M_\odot}\right)^{-5/3}\left(\frac{f_{\rm GW}}{0.01\,{\rm Hz}}\right)^{-8/3},
\label{eq:insptime}
\end{equation}
where $\mathcal M_{\rm c}=(1+z)(M_1M_2)^{3/5}(M_1+M_2)^{-1/5}$ is the redshifted chirp mass of the binary with component masses $M_1,\,M_2$, and $M_\odot$ is a solar mass. For solar-mass binaries detectable by LGWA, the redshift $z$ will be small and can be neglected. With a GW signal at $f_{\rm GW}=0.1\,$Hz, the inspiral time would be reduced to 3 years. 

We have seen that LGWA's response exhibits sharp peaks in the mHz band. A straight-forward estimation of the signal-to-noise ratio (SNR) of a solar-mass binary would be possible if the change in frequency of the signal over a few years observation time were significantly smaller than the width of these peaks. In this case, LGWA's response can be approximated as constant during the full observation (apart from a changing antenna pattern due to Moon's rotation). The time a binary spends on a specific normal mode can be calculated from equation (\ref{eq:insptime}) and is given by
\begin{equation}
\Delta T=122\,{\rm yr}\left(\frac{\mathcal M_{\rm c}}{2M_\odot}\right)^{-5/3}\left(\frac{f_n}{5\,{\rm mHz}}\right)^{-8/3}\frac{200}{Q_n},
\end{equation}
where the linewidth of a mode is $\Delta f_n=f_n/Q_n$. Certainly at 5\,mHz, the change in frequency is still so slow that one can approximate LGWA's response as constant over a few years even in the vicinity of a normal-mode resonance. This result is what allows us to use equation (\ref{eq:strainT}) to integrate the SNR of a GW signal over a few years. In real analyses, some necessary changes in signal frequency must be considered though, e.g., related to the orbital motion of the Moon causing Doppler shifts \cite{AsEA2014}. For signals at higher frequencies, the SNR must be integrated over a range of frequencies. 

Since a large range of masses will be considered in the following sections, inspiral times can vary greatly between different signals, which requires different sets of equations to estimate GW amplitudes and SNRs. In section \ref{sec:solar}, we estimate GW amplitudes of known solar-mass, compact binaries. For some of them, the orbital inclination angle is known, for others not. If the inclination angle is unknown, we use the inclination-angle and polarization averaged, time-domain GW amplitude
\begin{equation}
h = \sqrt{\frac{32}{5}}\frac{c}{r}(G\mathcal M_{\rm c}/c^3)^{5/3}(\pi f_{\rm GW})^{2/3},
\label{eq:averaged}
\end{equation}
where $c$ is the speed of light, $G$ the gravitational constant, and $r$ the luminosity distance of the GW source. The GW frequency $f_{\rm GW}$ is twice the orbital frequency of the binary. For compact binaries where an estimate of the inclination angle $\iota$ is available, we use the polarization average
\begin{equation}
\begin{split}
    h &= \sqrt{2(1+6\cos^2(\iota)+\cos^4(\iota))}\\
    &\qquad\cdot\frac{c}{r}(G\mathcal M_{\rm c}/c^3)^{5/3}(\pi f_{\rm GW})^{2/3}.
\end{split}
\label{eq:polarave}
\end{equation}
Some GW signals from compact binaries can evolve rapidly in frequency during the observation time as would be the case for DWDs above 0.1\,Hz or massive black-hole binaries throughout the entire LGWA band. In this case, a full signal spectrum needs to be calculated in the form of a Fourier-domain amplitude. Taking equation (\ref{eq:averaged}) together with results from Allen et al \cite{AlEA2012}, we obtain the inclination-angle and polarization averaged Fourier amplitude
\begin{equation}
|\tilde h(f)|=\sqrt{\frac{4\pi}{3}}\frac{c}{r}(G\mathcal M_{\rm c}/c^3)^{5/6}(\pi f)^{-7/6}
\label{eq:binarystrain}
\end{equation}
This equation is used in section \ref{sec:massive} to evaluate signal amplitudes of massive black-hole binaries. 

A convenient method to compare signal amplitudes with detector noise is to convert all spectra into characteristic strain. The characteristic strain $h_{\rm n}$ of detector noise is given by \cite{MCB2014}
\begin{equation}
    h_{\rm n}(f)=\sqrt{fS_n(f)},
\end{equation}
where $S_n(f)$ is the power spectral density of the detector noise. This needs to be confronted with the characteristic strain of GW signals. For signals with negligible frequency evolution, the characteristic strain can be written as
\begin{equation}
    h_{\rm c}(f)=\sqrt{2fT_{\rm obs}}\,h,
\end{equation}
where $h$ is the time-domain amplitude of the GW. Instead, if the signal is given as a Fourier spectrum $\tilde h(f)$, then the conversion into characteristic strain reads
\begin{equation}
    h_{\rm c}(f)=2f|\tilde h(f)|.
    \label{eq:charstrain}
\end{equation}
When confronting signal amplitudes with instrument noise in the following, we assume that only one seismometer monitors displacements, and that it is optimally located with respect to the propagation direction of the GW.

\subsubsection{Known stellar binary systems}
\label{sec:solar}
We collected from literature the orbital parameters (component masses, orbital period, orbital plane inclination) and estimated distance for a variety of known short-period binary systems in the Milky Way galaxy. With this information, we computed the expected characteristic strain $h_{\rm c}$ of the gravitational-wave emission based on equation (\ref{eq:polarave}). We scaled $h_{\rm c}$ to an assumed observing time T$_{\rm obs}=5$\,yr considering that all selected sources are monochromatic in this time interval. 

The list of binary systems includes:
\begin{itemize}
\item The LISA verification binaries from \cite{KuEA2018} and references therein. The list includes  11 semi-detached AM CVn type systems, 4 double WDs (DWD) and a WD with a hot He-rich companion. It has been suggested that the latter system  (empty green circle in figure \ref{fig:bs}) may evolve to become a type Ia SN \cite{geier2013} (single degenerate scenario, see below);
\item DWDs with known system parameters from the {\em ESO supernovae type Ia progenitor survey} (SPY) \cite{Napiwotzki2020} (double degenerate scenario); 
\item The DWDs discovered by the {\em Extremely Low Mass (ELM) Survey}, a spectroscopic survey targeting M$<0.3$ M$_\odot$ He-core WDs \cite{Brown2020};
\item The close binaries found by the {\em Systematic search of Zwicky Transient Facility data for ultracompact binary LISA-detectable GW source} \cite{BuEA2020}. These are mainly DWDs, but also AM CVn systems;
\item The ultra-compact X-ray binaries from \cite{Wen-cong2020}. UCXBs are low-mass X-ray binaries with ultra-short orbital periods made of a neutron star (NS) and a hydrogen-poor donor stars.
\end{itemize}

Location of the sources in the $h_{\rm c}$ vs. GW frequency plot are shown in figure \ref{fig:bs}. In the plot, we included the recurrent nova T-Pyxidis \cite{Patterson2017}, another representative of the single-degenerate (SD) scenario for SNIa, the high mass X-ray binary (HMXB) Cyg X-3 \cite{Koljonen2017} likely a BH+WR star, and two short period BNS PSR B1913+16 \cite{hulse1975}  and PSR J0737+3039 \cite{Burgay2003}.

The figure shows that up to a dozen known binary systems are bright enough in GWs to be detected by LGWA. UCXBs, while in a suitable frequency range, appear too faint for LGWA. HMXBs are brighter, but they never enter in the useful frequency range.

\begin{figure}
\includegraphics[width=0.95\columnwidth]{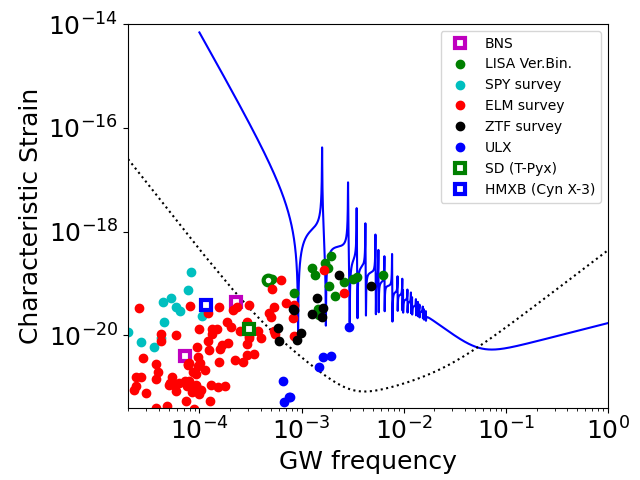}
\caption{The strength of the GW emission from known short-period binary systems in the Galaxy is compared with the expected sensitivity of LGWA (blue line, cryomagnetic) and LISA (black dotted line).}
\label{fig:bs}
\end{figure}

We acknowledge that the list of known binaries is severely biased. White dwarfs are electromagnetically faint and therefore they can be detected only at relatively short distances. In addition, DWDs can be identified only in favorable conditions, e.g. through the mutual eclipses when our line of sight lies close to the orbital plane. 

In fact, recent detailed simulations based on Galaxy structure modelling and binary population synthesis (e.g., \cite{Lamberts2019,Korol2019}) predicted that LISA will be able to resolve signals for up to $10^5$ compact binaries of which only a small fraction ($\sim 10^2$ events) will also be identified by EM radiation. 

For LGWA, we propose a similar kind of analysis when an accurate estimate of the array sensitivity will be available. Here, we perform a simple back of the envelope calculation to test the possible performances for a specific science case, which is to identify the progenitors of type SN~Ia.

\subsubsection{The case of SN Ia progenitors}

SN~Ia have a crucial role in cosmology for the measurement of the Hubble constant \cite{Riess2019,Khetan2020} and the discovery of cosmic expansion acceleration \cite{Riess1998,Perlmutter1999}. Yet, it is still unclear what is the evolutionary path that leads to explosion \cite{mdv2020}. The standard scenario calls for a thermonuclear explosion of a WD that grows above the Chandrasekhar mass limit. However, there are two different scenarios that can plausibly lead to this event, both involving compact binaries:  a) the merging of two WDs (double degenerate scenario) or b) WD mass accretion from a non-degenerate companion, either a main sequence or a red giant star (single degenerate scenario).

For double degenerate systems, merging occurs when the system loses angular momentum due to GW emission while for single degenerate systems the time scale is set by the stellar evolution clock of the secondary \cite{Greggio2010}. This implies a different distribution of orbital periods for the two scenarios. 

In the double degenerate scenario, the distribution of the DWD orbital periods can be derived knowing the merging time from GW emission using equation (\ref{eq:insptime}). The total number of DWDs in a selected orbital period range (note that only systems with  total mass $>1.4$ M$_\odot$ are relevant) is constrained by the current rate of SN~Ia in the Galaxy:  for the current estimate of 5 SN~Ia for 1000\,yr \cite{Li2011}, the total number of DWDs with orbital period below $5\times10^3$\,s needs to be $\sim 240,000$. Exploiting the fact that the SN~Ia rate in the Galaxy is almost constant for time scale of a billion year \cite{Greggio2020}, the orbital period distribution can be easily derived assuming a uniform distribution of the merging time.

For the estimate of the expected GW signal distribution we need to know the distance distribution of the DWD systems. For this exploratory calculation, we take that the systems are distributed as the stars in the Galactic disk modeled with an exponential profile with scale $R_0=2.5$\,kpc. 

With these assumptions, we performed a simple Monte Carlo experiment selecting random orbital periods and random distances following the adopted distributions. For each system, we computed $h_{\rm c}$, the GW characteristic strain, and then selected the systems with $h_{\rm c}$ higher than the expected array sensitivity $h_{\rm n}$. We also selected the sub-sample of systems with $h_{\rm c}/h_{\rm n}>7$ that we adopt as the limit for resolved sources.  

\begin{figure}[ht]
\includegraphics[width=0.95\columnwidth]{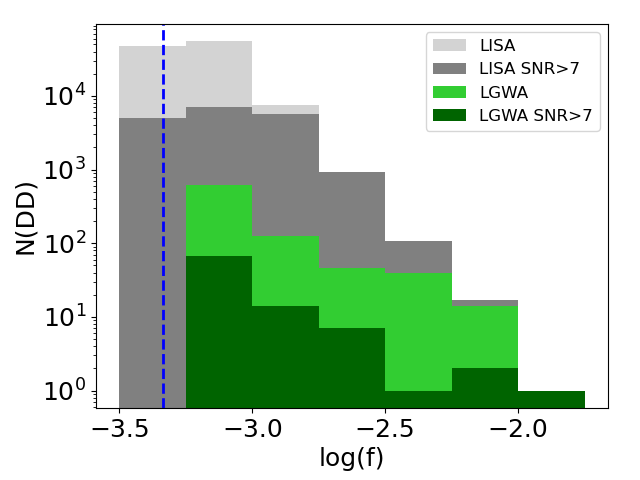}
\caption{Monte Carlo simulation of the GW frequency distribution for double degenerate (DD) systems in the Galaxy detectable by LGWA and LISA with different detection threshold.
The blue dashed line indicates the expected maximum GW frequency (corresponding to the minimum orbital period) for SD systems.}
\label{fig:dd}
\end{figure}

The frequency histogram for the detected sources is shown in figure \ref{fig:dd}. We concluded that, if double degenerate is the only viable path to SN~Ia, LISA should be able to resolve $\sim 10^3$ DWDs with total mass $>1.4$ M$_\odot$  while LGWA is expected to detect $\sim 10^2$ DWDs. Considering that the minimum period for the single-degenerate scenario is 1.2\,hours  \cite{Davis2010,Patterson2017}), few, if any, of these system can be detected with $\log(f)>-3.3$ and even LGWA  would then be able to give a definite answer to the long standing issue of Ia progenitors.

\begin{figure}[ht]
\includegraphics[width=0.95\columnwidth]{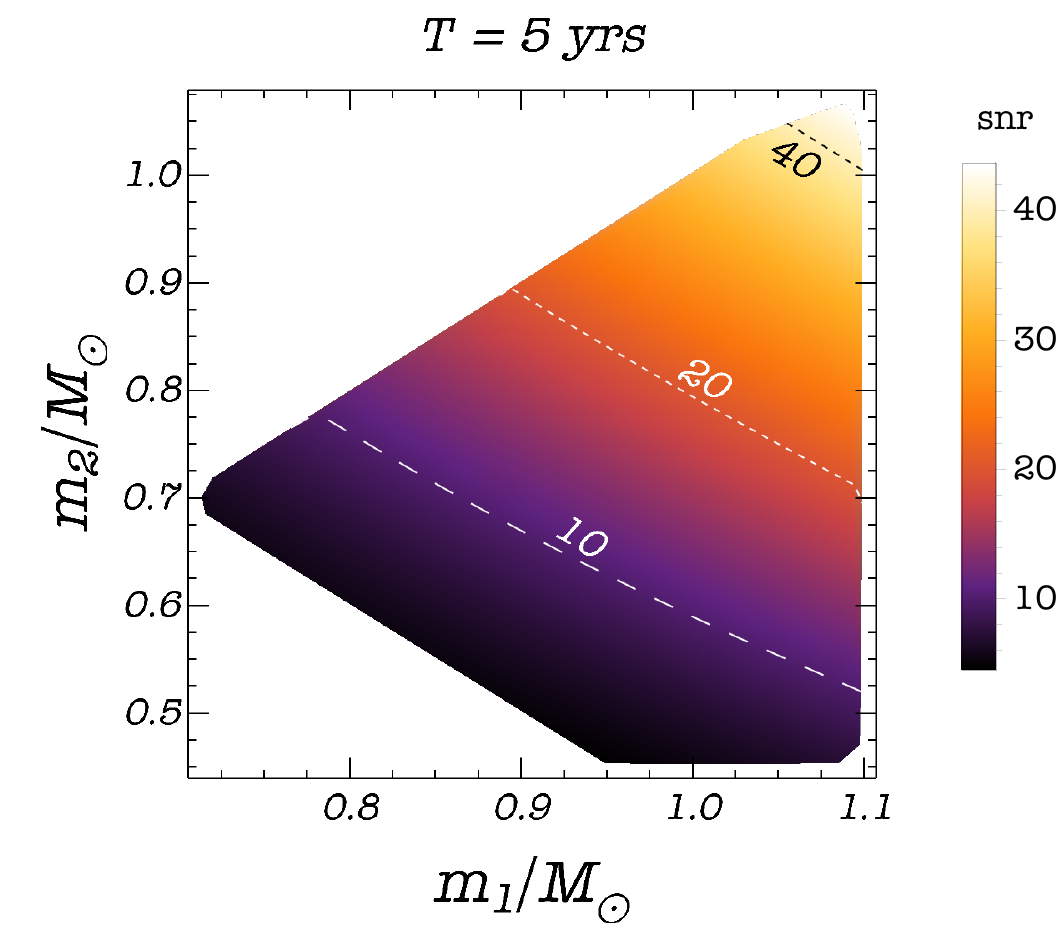}
\caption{SNR as a function of the component masses of the 
binary system of WD at the distance of 1 Mpc observed for 5 
years before the merger. Dashed curves correspond configurations 
of constant SNR. The WD masses are chosen to fall within the 
range where the coalescence may lead to a supernovae Ia event 
\cite{Postnov:2014tza}.}
\label{fig:WDmerger}
\end{figure}

LGWA's observation band covers the merger frequencies of DWDs. Figure \ref{fig:WDmerger} shows the SNR as a function of the component masses of the binary system at the distance of 1\,Mpc observed for 5 years before the merger. Considering an SNR threshold of 10 for the detection, the merger of the most massive systems can be observed up to 4--5\,Mpc. The rate of SN~Ia up to this distances is about 0.02 per year~\cite{Li2011}. Even if the probability of a joint detection is low, detecting a GW signal from the DWD merger together with the optical or high energy emission of a SN~Ia has an enormous scientific impact to shed light on the progenitor of SN~Ia and for cosmological studies~\cite[e.g.][]{Maselli2020}. Furthermore, the probability of a joint detection can increase using SN~Ia exploded within ten Mpc to drive the GW search of LGWA data as done using gamma-ray bursts or optical core-collapse SNe for the LIGO and Virgo data~\citep[e.g.][]{Abbott2020}.

A prompt association of GW signals from double degenerate systems with SN~Ia could be achieved by the coincident detection of short-lasting X-ray/gamma-ray signals from a shock breakout \citep{Colgate1974} (SBO). The characteristics of SBO signals, i.e., duration, temperature and luminosity strongly depend on the fundamental properties of an explosion including the size of the progenitor (or an extended region such as wind or debris), its mass, and the total energy of an explosion \citep{Katz2010,Piro2010,Budnik2010,NakarSari2010,NakarSari2012}. Particularly for SN~Ia, it is yet unclear whether the SBO is relativistic or Newtonian, since the velocity of the shock front strongly depends on the radius of a breakout. In a single degenerate model, this radius is expected to coincide with the size of a WD $\sim 1000$\,km, which would produce a 10\,ms flash of low-luminosity  $\sim 10^{44}$\,erg/s MeV photons by relativistic SBO \citep{NakarSari2012}. In a double degenerate model, the optical depth of the progenitor surroundings could cause a delay of the SBO at $\sim 10^{11}$- $10^{12}$\,m, which corresponds to a Newtonian or sub-relativistic SBO \citep{Fryer2010}. In this case, the expected signal would be much longer $\sim 10^3$\,s, brighter $\sim 10^{47}$\,erg/s and at softer energy bands ($\sim$ 50--200\,keV) \citep{NakarSari2010}. The brief initial flash of SBO is followed by its fainter cooling tail which constantly softens and can be observed in soft X-rays and in the UV range. With respect to the optical band, the advantages of detecting the SN~Ia through the SBO are the shorter delay between the GW signal and the electromagnetic signal, and the fact that the high-energy emission does not suffer from dust absorption. Large field-of-view satellites observing in the X-rays, such as Einstein Probe~\cite{EP2015}, SVOM-ECLAIRs~\cite{SVOM2016}, and the mission concept THESEUS~\cite{Amati2018} are expected to operate in the next decade. 

\subsubsection{Massive and supermassive black-hole binaries}
\label{sec:massive}

There is a clear consensus from the observations that the majority of massive galaxies should harbour a central massive black hole (MBH, with masses from few times $10^5$ to $10^{10} M_\odot$), whose evolution is deeply intertwined with their host \citep{kormendy95,ferrarese00,haring04,kormendy13}. Following the current $\Lambda$CDM cosmological paradigm, galaxies grow hierarchically through minor and major mergers \citep{springel05,hopkins05}. As these black holes share a symbiotic relation with their host galaxy, they should play a significant role during galaxy mergers. Indeed, during a galaxy encounter, the dark matter, gas and stars of both galaxies form a common envelope around the two black holes, which undergo a process called dynamical friction \citep[e.g.,][]{sesana10,dosopoulou17}. Here the interactions between the black holes and the surrounding matter drag the two black holes towards the center of the new common galactic envelope. Eventually, the two black holes will bind in a binary system with sub-parsec scale distance \citep[e.g.,][]{begelman80} and emit continuous gravitational waves in the frequency range  $f\sim1-100\;\text{nHz}$. Despite the significant difficulties to observe such kind of binary systems, several candidates have been proposed by using both optical \citep[e.g.,][]{graham15a,graham15b,charisi16} and X-ray observations \citep{severgnini18,serafinelli20}. The emission of gravitational waves from the binary will effect the binary period, shrinking their distance at every cycle until they eventually merge \citep[e.g.,][]{dotti12,mayer13,colpi14}.
\begin{figure}[ht]
\centering
\includegraphics[width=1.1\columnwidth]{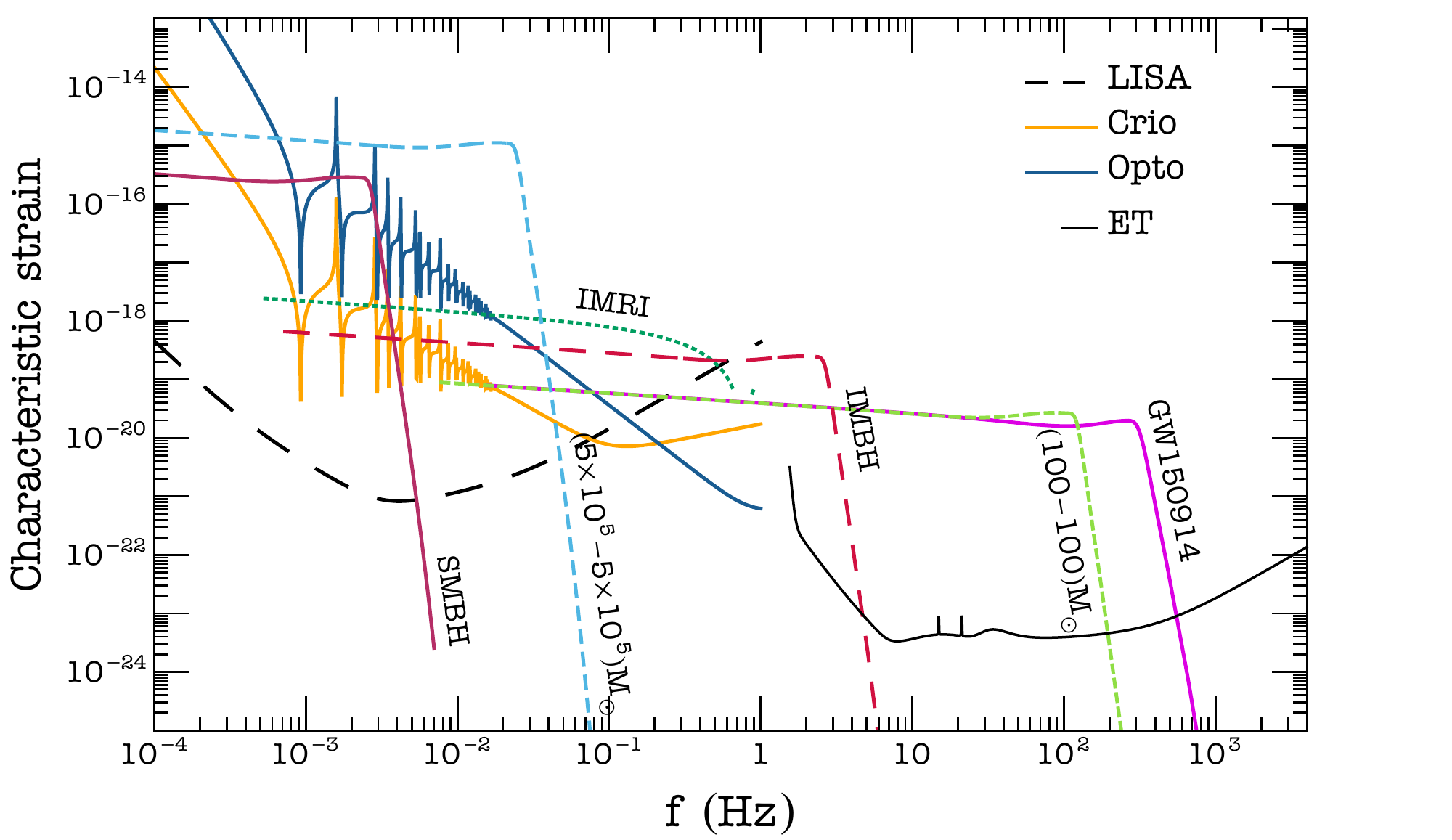}
\caption{Characteristic strain for various sources. The amplitude is modelled using the phenomD template in the frequency domain, including post-Newtonian corrections \cite{Khan:2015jqa,Husa:2015iqa}. The SMBH system is positioned at a luminosity distance of $\sim 1.7$\,Gpc (SNR = 500), the IMBH at a distance of $\sim 2$\,Gpc (SNR = 50), the IMRI at $\sim830$\,Mpc  (SNR = 100). The GW150914-like system and the (100--100)$M_\odot$ are positioned at a luminosity distance corresponding to an SNR = 10, i.e., $\sim160$\,Mpc and $\sim412$\,Mpc, respectively.}
\label{fig:strain}
\end{figure}

\begin{figure}[ht]
\includegraphics[width=0.95\columnwidth]{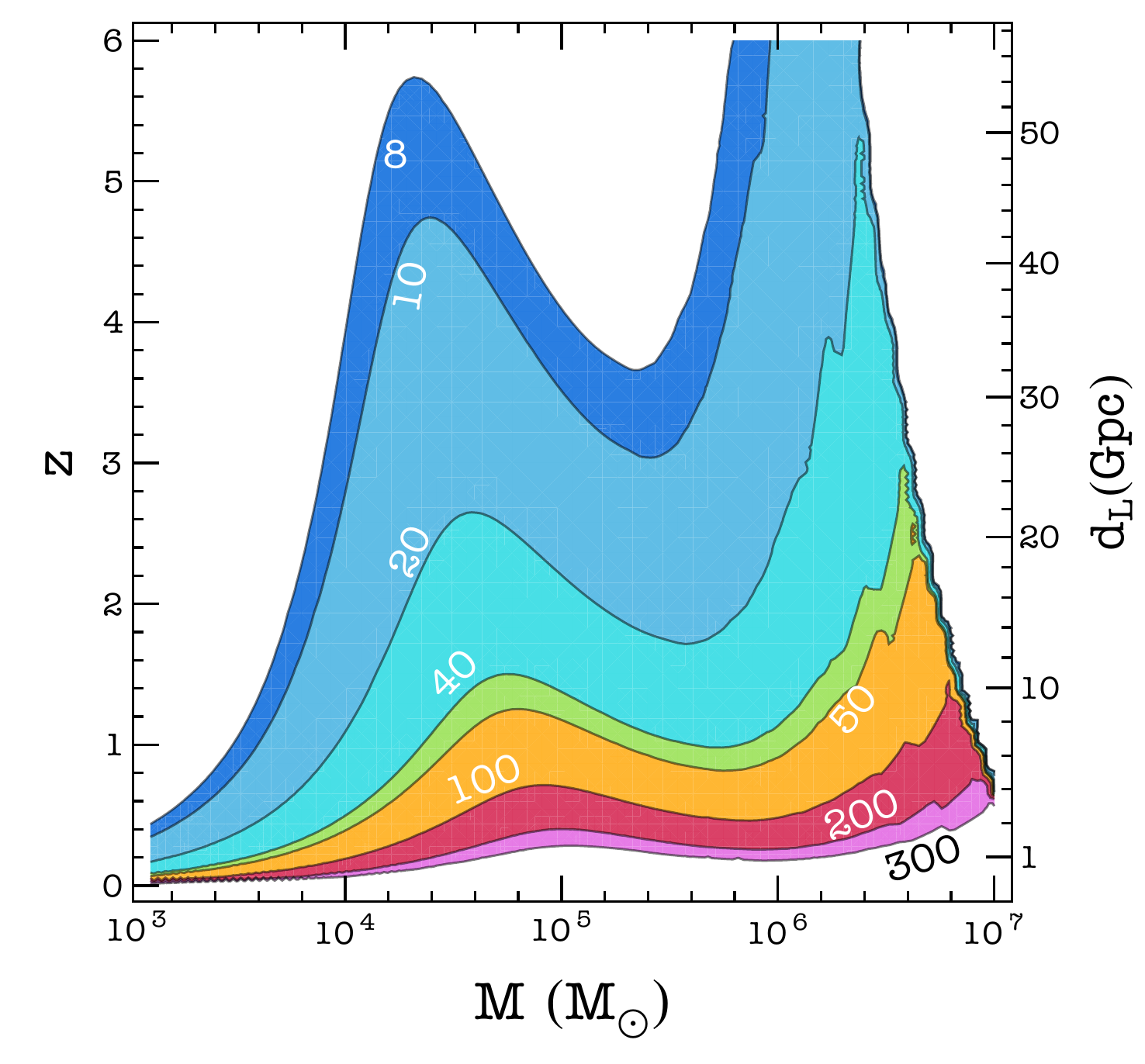}
\caption{Coalescences of massive and supermassive black-hole binary. 
The contours of constant SNR are shown as function of total source-frame 
mass, $M$, and of redshift, $z$ (left axis) and luminosity distance, 
$d_\textnormal{L}$ (right axis), considering binaries with constant 
mass ratio among the component masses of 0.2.}
\label{fig:BBHSNR}
\end{figure}

The gravitational event from the orbital, merger and ringdown of the two black holes is in principle observable by low-frequency gravitational wave observatories. To assess the observability of these sources with LGWA, we produced waveforms of several binaries (see figure \ref{fig:strain}) using the PhenomB template (see section \ref{sec:fundamental} for details). A binary system with $M_1=5\times10^6M_\odot$ and $M_2=4\times10^6M_\odot$ at a distance of $d\sim2$ Gpc, and a closer system with $M_1=M_2=5\times 10^5M_\odot$ at $d=50$ Mpc are both shown in figure \ref{fig:strain}. Boh these systems have extremeley high signal-to-noise ratio (SNR$\gg10$), and therefore LGWA will be able to follow part of the orbital phase, as well as the merger and ringdown of a coalescence in the mass range $M\lesssim10^7M_\odot$, even at large redshift.
In fact, considering a detection threshold of SNR $\sim10$, a system with $M_1=M_2=5\times10^5M_\odot$ can be observed up to $z\simeq3$. 
Figure \ref{fig:BBHSNR} shows the SNR as a function of the source-frame total mass of 
these binaries and of the redshift (or of the luminosity 
distance following \cite{ AdEA2016}).

\indent It is also very interesting to discuss the implications for the EM observations of massive binaries of a possible LGWA detection, and vice versa. This is particularly relevant, because LGWA will have a sky localization, allowing us to point at the emitting source right after the GW detection. The only known type of astrophysical objects powered by a MBH are active galactic nuclei (AGN). AGN are powered by the accretion of matter in a disk around a MBH. Such accretion disk is responsible for the optical and ultraviolet (UV) emissions, that may exceed the one emitted by the entire host galaxy. Additionally, a hot electron plasma in the region closest to the black hole, the so-called corona, up-Comptonizes the UV photons, emitting X-rays \citep{haardt91,haardt93}. When there is a massive black-hole binary system, the physics related to the accretion disk becomes more complex. In fact, simulations suggest that the two black holes will excavate a cavity inside a circumbinary disk, and the accretion will occur by means of two minidisks, surrounding each of the black holes \citep[e.g,][]{hayasaki08,noble12,dascoli18}. While the circumbinary disk is responsible for the optical emission, UV photons and X-rays are likely emitted by the minidisks and the coronas; therefore X-rays are one of the most promising tools to identify binary systems by means of their EM signal. Before the coalescence, as these emissions are produced in the minidisks, a periodic modulation in the UV and X-ray light curves might be expected. An additional feature that may arise is a  double peaked  Fe K$\alpha$ emission line. This Fe emission line is the superposition of the two Fe emission lines arising from each of the mini disks, which are Doppler shifted due to the orbital motion \citep[e.g.,][]{popovic12,sesana12,roedig14,mckernan15,farris15,haiman17}. Finding these features in the X-rays would be crucial to assess the abundance of MBH merging events in the Universe, and, possibly, they would allow us to predict low-frequency black-hole merger GW signals. In fact, simulations show that this configuration holds until a few orbits before the coalescence \citep{tang18}. The EM counterparts of both the merging and ringdown phases are currently unknown and represent an exciting challenge of low-frequency gravitational-wave astronomy.

\subsection{Spinning neutron stars}

The latest releases (v 1.64) of the ATNF (Australia Telescope National Facility) catalog \cite{manch+05}, and of the McGill Online Magnetar catalog \cite{ok14} contain in total 1045 neutron stars (NSs) spinning at a period in the [0.04-1.5]\,Hz range. Most of them (1021) are ``ordinary" radio pulsars, the rotational pace of which can be easily monitored in the electromagnetic band by using the procedure of the pulsar timing \citep[e.g.,][]{poss+burg16}. Another 24 objects belong to the category of the so-called magnetars \cite{mereg+15}, the rotational parameters of which are mostly determined via timing observations in the X-ray band.  At the present level of knowledge, none of these NSs is included in a binary system. 

All these NSs undergo a secular spin down, which is commonly believed to mostly arise from the conversion of rotational energy (for the ordinary radio pulsars) and of magnetic energy (for the magnetars) into radiation and particle acceleration.

However, if the NS possesses a nonzero mass quadrupole moment $Q_{22}$, and assuming polarization properties in agreement with the predictions of general relativity, the rotation of the NS also generates the emission of GWs, which might be partly responsible for the total energy loss, and hence playing a role in the observed slow-down of these objects \citep[e.g.,][]{zimm+szed79}. In particular, in the simplest picture, the GWs are emitted at a frequency $f_{\rm gw}=2f_{\rm sp}$, where $f_{\rm sp}$ is the spin frequency of the NS. On a theoretical ground, the detailed GW frequency spectrum could be more complex, including also emission at the same rotation rate of the star, as well as a small modulation about both the mentioned frequencies if the NS undergoes free precession \citep[e.g.,][]{zimm+szed79, jones+ander02}. However, the strongest GW signal is expected to be the one at twice $f_{\rm sp}$ \citep[e.g.,][]{jones15}, and we will only focus on that for the purpose of this analysis. 

In all cases, the fractional variations in the spin rate of these objects is small over a decadal timescale ($10^{-9}-10^{-3}$ for ordinary radio pulsars, $10^{-7}-3\times 10^{-2}$ for magnetars). Once the barycentric corrections are applied (i.e., the effects of the combined rotational and orbital motions of the Moon are removed), the expected GW emission can then be considered quasi-monochromatic. Even if the slow down of some of these sources can experience irregularities (e.g., glitches and rotational noise \citep[e.g.,][]{Lyne+gs12}), dedicated electromagnetic campaigns are able to compensate for that and suitably follow the rotational phase for the majority of the ordinary radio pulsars. In practice, the availability of electromagnetic information provides the possibility to integrate the GW signal associated with those NSs over multi-year long intervals. This holds true also for the population of the bright steady emitting magnetars. Only for the subclass of the transient magnetars, the X-ray luminosity during the quiescent phase can go below the threshold for monitoring their rotational behavior. In this case their putative GW signal could be optimally integrated for the duration of the bright phases only, typically lasting of order months. 
  
All the considerations above suggest to investigate if any potential targets for LGWA could be found in the mentioned populations of NSs.

In absence of any precession, the amplitude $h_0$ of the GWs released by a triaxial NS spinning around a principal axis (assumed to be the $z$ axis), due to a mass quadrupole $Q_{22}\neq 0,$ is \citep[e.g.,][]{aasi+++14}
\begin{equation}
  h_0 = \frac{16\pi^2G}{c^4}\frac{I_{zz}f_{\rm sp}^2\epsilon}{d}\simeq4.2\times 10^{-30} d_{\rm kpc}^{-1} I_{zz,45} f_{\rm sp,1}^2 \epsilon_{-6},
\end{equation}
where, for the sample of NSs under consideration, the distance $d_{\rm kpc}$ in units of kpc lays in the range 0.11 -- 59.7 and is measured with typical maximum uncertainty of about 50\%. The spin frequency in Hz, $f_{\rm sp,1},$ is in the range 0.04 -- 1.5 and is measured with negligible uncertainty with respect to the other parameters in the equation above. The value of the component $I_{zz}$ along the spin axis of the moment-of-inertia tensor ellipsoid results from theoretical calculation: although it is dependent on the NS mass and the adopted equation of state, the predicted values $I_{zz,45}$ span a small range of about 1 -- 3 when expressed in units of $10^{45}$ g cm$^2.$    

The most uncertain parameter is the ellipticity $\epsilon$ (expressed in units of $10^{-6}$ in the equation above), which is defined on the basis of the true moments of inertia $I_{xx}$ and $I_{yy}$ about the other two principal axes $x$ and $y$ (left part of the equation below), and which also relates to the mass quadrupole moment $Q_{22}$ according to the right part of the following equation \citep[e.g.,][]{aasi+++14}:
\begin{equation}
  \frac{|I_{xx}-I_{yy}|}{I_{zz}} \equiv \epsilon = \sqrt{\frac{8\pi}{15}}\frac{Q_{22}}{I_{zz}}  
\end{equation}
The Advanced LIGO (aLIGO) experiment has investigated the case of 221 NSs, searching for GW signals at frequencies above 10\,Hz \cite{LigoPsr19,LigoPsr19_err}. The lack of any detection makes it possible to constrain the value of the ellipticity for that sample of rapidly spinning pulsars. The smallest observed upper limit is $\epsilon_{-6}\sim 6 \times 10^{-3},$ measured in a pulsar spinning at about 350\,Hz, in the frequency interval for which aLIGO had the best sensitivity. The upper limits to the ellipticity scale linearly with those on the GW amplitude (i.e., with the sensitivity at a given frequency, which gets worse below 100\,Hz in the case of the aLIGO experiment), as well as with $f_{\rm sp}^{-2}$.  Therefore, when looking at slower spinning targets, upper limits on $\epsilon_{-6}$ becomes much less constraining, i.e. 1 -- 10$^3$. For instance Abbott {\it et al.} \cite{LigoPsr19} found $\epsilon_{-6}<10$ and $\epsilon_{-6}<76$ for the Crab and the Vela pulsar, respectively.
 
\begin{figure}
\centering
\vspace{-2.5truecm}
\includegraphics[scale=0.4]{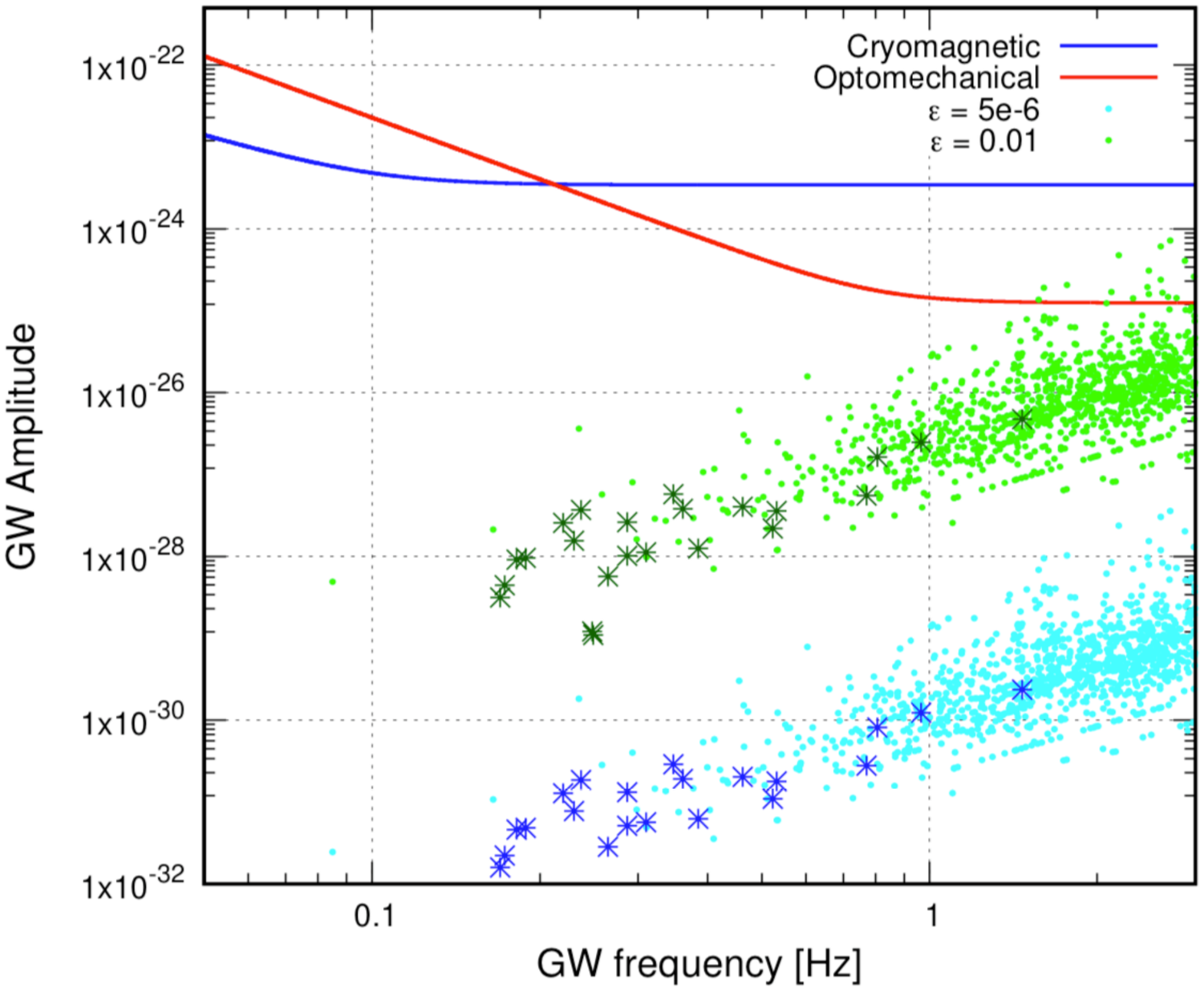}
\vspace{-2.5truecm}
\caption{Amplitude of the expected GW signals vs the GW frequency for the sample of the NSs listed in the version 1.64 (Oct 2020) of the ATNF catalog ({\small {\tt https://www.atnf.csiro.au/people/pulsar/psrcat/}}) or in the McGill Online Magnetar catalog ({\tt http://www.physics.mcgill.ca/$\sim$pulsar/magnetar/main}). All the reported NSs have a known distance and spin at rotational frequencies in the range 0.04--1.5\,Hz. Two values for the ellipticity $\epsilon$ (expressed in units of $10^{-6}$) have been applied: $\epsilon_{-6}=5$ ({\it cyan dots} and {\it blue stars} ) and $\epsilon_{-6}=10^4$ ({\it green dots} and {\it dark green stars}). The value of the component of the moment-of-inertia tensor ellipsoid along the spin axis $I_{zz,45}$ is set equal to 1 in units of $10^{45}$\,g cm$^2$ (see text). In particular the sample of the so-called "magnetars" is represented with {\it stars}, whereas {\it circles} are associated with the other NSs in the sample. The predicted sensitivity curves of LGWA are also reported for two different designs of the detector: Optomechanical ({\it solid red line}) and Cryomagnetic ({\it solid blue line}). They are calculated assuming an integration time of 5 years and a minimal signal-to-noise ratio for a fiducial detection equal to 5. }
\label{fig:strainSpinNS}
\end{figure}

The latter constraints are close to the theoretical predictions (suggesting $\epsilon_{-6}^{\rm max}$ in the interval 1 -- 10) about the maximum values of the ellipticity associated with a NS of mass 1.2 -- 2.0 $M_{\rm \odot}$ for a wide range of equations of state \cite{johnOwen13} and in absence of a significant magnetic energy stored inside the NS. Hence in figure \ref{fig:strainSpinNS}, we adopted a fiducial maximum value $\epsilon_{-6}^{\rm max}=5$ in order to plot the most optimistic GW amplitudes that we can get from our sample of spinning NSs, supposing that they do not contain ultra-strong magnetic fields, nor exotic particles. Those amplitudes are compared with the estimated sensitivity of LGWA to quasi-monochromatic GW signals, assuming an integration time of 5 yrs and a signal-to-noise threshold SNR$_0=5$ (see eq \ref{eq:strainT}).  
   
None of the selected targets seems suitable to be detected by LGWA, under the aforementioned hypothesis. We also note that, the adopted value of $\epsilon_{-6}^{\rm max}$ is maybe even too optimistic on an astrophysical ground, in the absence of a strong magnetic field inside the NS. In fact, using a hierarchical Bayesian approach, Pitkin {\it et al.} \cite{pitkin+18} combined data from individual sources observed during LIGO’s S6 run in order to derive an upper limit $\sim 5\times 10^{-2}$ for the mean value of the distribution of $\epsilon_{-6},$ suggesting that (at least the bulk of) the pulsar population is not strained to its maximum possible value of the ellipticity. 

Perspectives for a detection with LGWA might significantly improve if one accounts for the possible role played by a strong magnetic field buried in the interiors of some of the observed NSs \cite{DaPe2017,LaJo2018}. Frieben \& Rezzolla \citep[]{f+rezz12} calculated that, for a favorable magnetic field topology, internal toroidal magnetic fields of order $10^{15}$\,G (in the ball-park of the observed values of $B_{\rm p},$ the external magnetic field measured at the pole of the ``magnetars'') are able to deform a typical 1.4 M$_\odot$ NS to ellipticity values reaching $\epsilon_{-6}=100$.  Moreover, when the internal toroidal component dominates the total magnetic energy budget of the NS, the values of $\epsilon_{-6}$ scales as $B_{\rm p}^2$ \cite{c+rezz13}, in principle generating magnetic deformations of the NS up to $\epsilon_{-6}=10^3$ for internal magnetic fields of order $10^{16}$\,G and up to $\epsilon_{-6}=10^5$ for the maximum allowed B-field strength $\sim 10^{17}$\,G. 
    
Large values for the ellipticity in condition of maximum strain $\epsilon_{-6}^{\rm max}\gapp 10^3$ can also be approached if some NSs are indeed containing exotic particles \cite{johnOwen13}, like is the case for the putative ``hybrid'' stars (i.e., stars having a hadron-quark mixed phase lattice in their core \citep[e.g.,][]{owen05}), and $\epsilon_{-6}^{\rm max}\sim 10^5$ can be sustained by elastic deformations in putative solid quark stars \cite{johnOwen13}.

In view of these theoretical investigations, we also report in figure \ref{fig:strainSpinNS} the expected GW amplitudes for our sample in case $\epsilon_{-6}=10^4$. In this case there appears a handful of potential targets, which could be representative of a small subgroup among the observed NSs, which are endowed with huge internal magnetic field and/or comprise exotic particles. Despite this admittedly being the most favorable predictable situation, it is interesting to note that LGWA would be the sole instrument capable of discovering such intriguing objects in the mentioned frequency range, since LISA is not expected to have the required sensitivity. 

\subsection{Fundamental physics}
\label{sec:fundamental}
The low-frequency/long-wavelength GW signals observed by LWGA 
represent a promising opportunity to study fundamental physics. 
LGWA is expected to detect sources in a wide range of masses, 
which will allow to test the nature of gravity in the strong 
field regime, and to challenge general-relativity (GR) predictions \cite{Berti:2015itd}. 
Observations by LGWA will be independent and complementary to 
those obtained by current and future ground-based 
interferometers \cite{Sathyaprakash:2019yqt,Perkins:2020tra}, leading to searches 
of possible hints of new physics in a different class of sources. 
Some of the science cases that will benefit from LGWA observations 
include: (i) tests of non-GR signatures within the GW generation and 
propagation mechanisms, (ii) tests of the nature of black holes, and 
of the existence of new degrees of freedom which modify the 
gravitational interaction and/or of exotic states of matter, (iii) 
probes of dark matter and of dark energy models, (iv) characterization 
of the environment in which compact binaries evolve and merge. 
We refer the reader to
\cite{Yunes:2010qb,Berti:2018cxi,Berti:2018vdi,Barack:2018yly,Berti:2019xgr,Cardoso:2019rvt} and reference therein for a detailed discussion 
on these and further fundamental physics topics relevant for the frequency range in which LGWA is sensitive. Hereafter instead, we will focus on two specific classes of tests, studying the actual constraints that 
LGWA will be able to infer on GR modifications, based on the 
cryomagnetic noise spectral density described in the previous sections.

As a first science case we focus on model independent tests of 
gravity in the inspiral part of the signal emitted by compact 
binaries. Theory agnostic approaches allow to study possible 
GR modifications, testing the validity of the post-Newtonaian 
structure of the GW signals. We consider here the parametrized 
post Einsteinian framework (ppE) \cite{Yunes:2009ke}, that introduces 
shifts within both the amplitude and the phase of the waveform, 
which can be measured using GW data. Constraints on 
such parameters can be mapped to bounds on the fundamental 
couplings of specific theories of gravity, alternative to GR 
\cite{Yunes:2013dva}. This approach roots into the well known 
parameterized post-Newtonian (ppN) formalism developed by Will and Nordtvedt \cite{1971ApJ...169..125W,1971ApJ...163..611W,1972ApJ...177..775N,1972ApJ...177..757W}, 
and it is analogous to the parametric analysis routinely 
performed by the Virgo/LIGO collaboration \cite{AbEA2019b,AbEA2016c}.

\begin{figure}[ht]
\centering
\includegraphics[width=0.95\columnwidth]{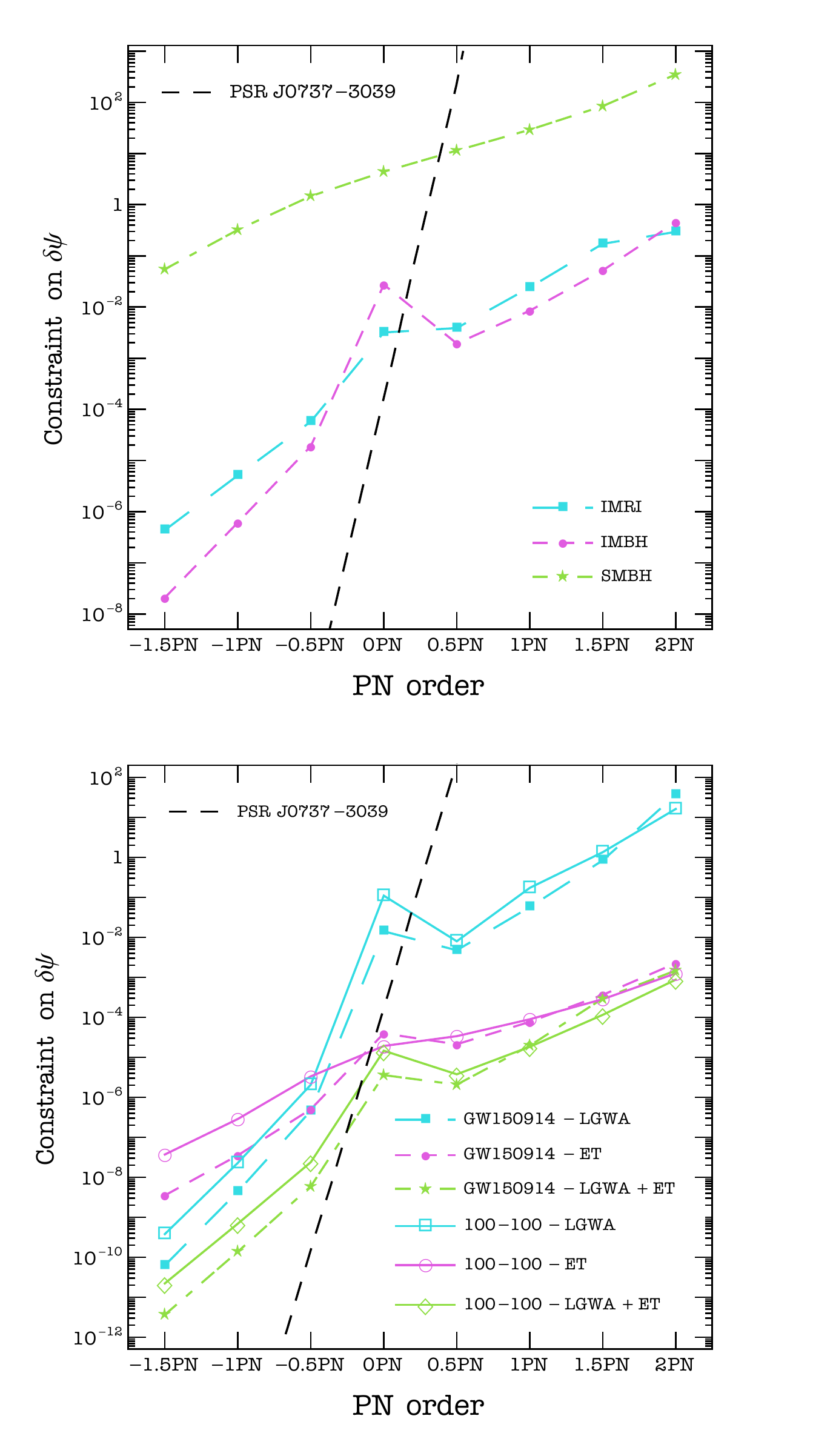}
\caption{(Top) Constraints on agnostic deviations from GR in 
the inspiral phase of different classes of compact binaries. 
The horizontal axis shows the post-Newtonian order of the 
waveform modification. The dashed black line identifies the agnostic 
constraints inferred form binary pulsar observations 
\cite{Yunes:2010qb}. Regions above the curves are excluded 
by observations. (Bottom) Same as top panel, but for a 
a system with the same masses and spins of GW150914, and 
a binary with $m_1=m_2=100M_\odot$, observed by 
LGWA and by a 3G ground-based detector like the Telescope. 
Signals are integrated for 5 years before the end of the 
inspiral phase, as defined in \cite{Ajith:2007kx}, and the 
luminosity distance is fixed to have an SNR of 10 in LGWA.}
\label{fig:ppetest1}
\end{figure}
The top panel of figure \ref{fig:ppetest1} shows the projected 
constraints obtained by LGWA using a Fisher matrix analysis
\cite{Sathyaprakash:2009xs},  for modifications of the GW phase 
at different PN orders, for three type of sources: (i) intermediate 
mass ratio inspirals (IMRI) with masses $(m_1,m_2)=(10^5,10^3)M_\odot$, 
(ii) intermediate mass BHs (IMBH) featuring $(m_1,m_2)=(5,4)\times10^3M_\odot$, 
and (iii) supermassive BH binaries (SMBH) with $(m_1,m_2)=(5,4)\times10^6M_\odot$.
The GW signal is modelled through the PhenomB template in the frequency 
domain for non-precessing spinning BHs \cite{Ajith:2007kx,Ajith:2009bn} 
augmeted by the ppE phase shift, with Newtonian amplitude averaged on the source orientation, and distance set in order to have a SNR equal to 100, 50 and 500, for IMRI, IMBH and SMBH, respectively, which correspond to a luminosity distance of 
$d\simeq (827,1954,1698)$Mpc. 
For all binaries the dimensionless spins 
parameters are fixed to $(\chi_1,\chi_2)=(0.7,0.9)$. The bounds 
scale as the inverse of the SNR. Therefore, the values shown in 
figure \ref{fig:ppetest1} can be easily rescaled to any value of the SNR.

For comparison, we also show similar constraints inferred from 
binary pulsars observations in the electromagnetic band 
\cite{Yunes:2010qb}, with the region above each curve 
being ruled out by observations. For PN coefficients $\gtrsim 0$, 
projected constraints by LGWA will be stronger than those obtained 
by pulsars. For negative PN corrections which are more relevant for 
systems with large orbital separation, i.e. small frequencies, 
LGWA will be also able to derive strong bounds on the GW phase 
shifts, although those coming from binary pulsars will still 
be dominant. However, it is worth remarking that the single-case 
scenario analysed here will also benefit from multiple GW 
observations, which can be easily combined into this approach 
in order to further lower the bounds on the phase shifts. 
For low mass sources, LGWA can also be exploited to perform 
multi-band test of GR. The bottom panel of figure \ref{fig:ppetest1} 
shows indeed the constraints on a (i) GW150914-like system and 
(ii) a $(100-100)M_\odot$ binary with $(\chi_1,\chi_2)=(0.7,0.9)$, 
observed both by LGWA and by a 3G detector like Einstein 
Telescope (ET) \cite{PuEA2010}. This analysis shows how 
the synergy between LGWA and future ground based detectors 
can narrow down the bounds on the GR deviations on a wide 
range of PN orders, active at both lower and high 
frequencies. 

\begin{figure}[ht]
\centering
\includegraphics[width=0.95\columnwidth]{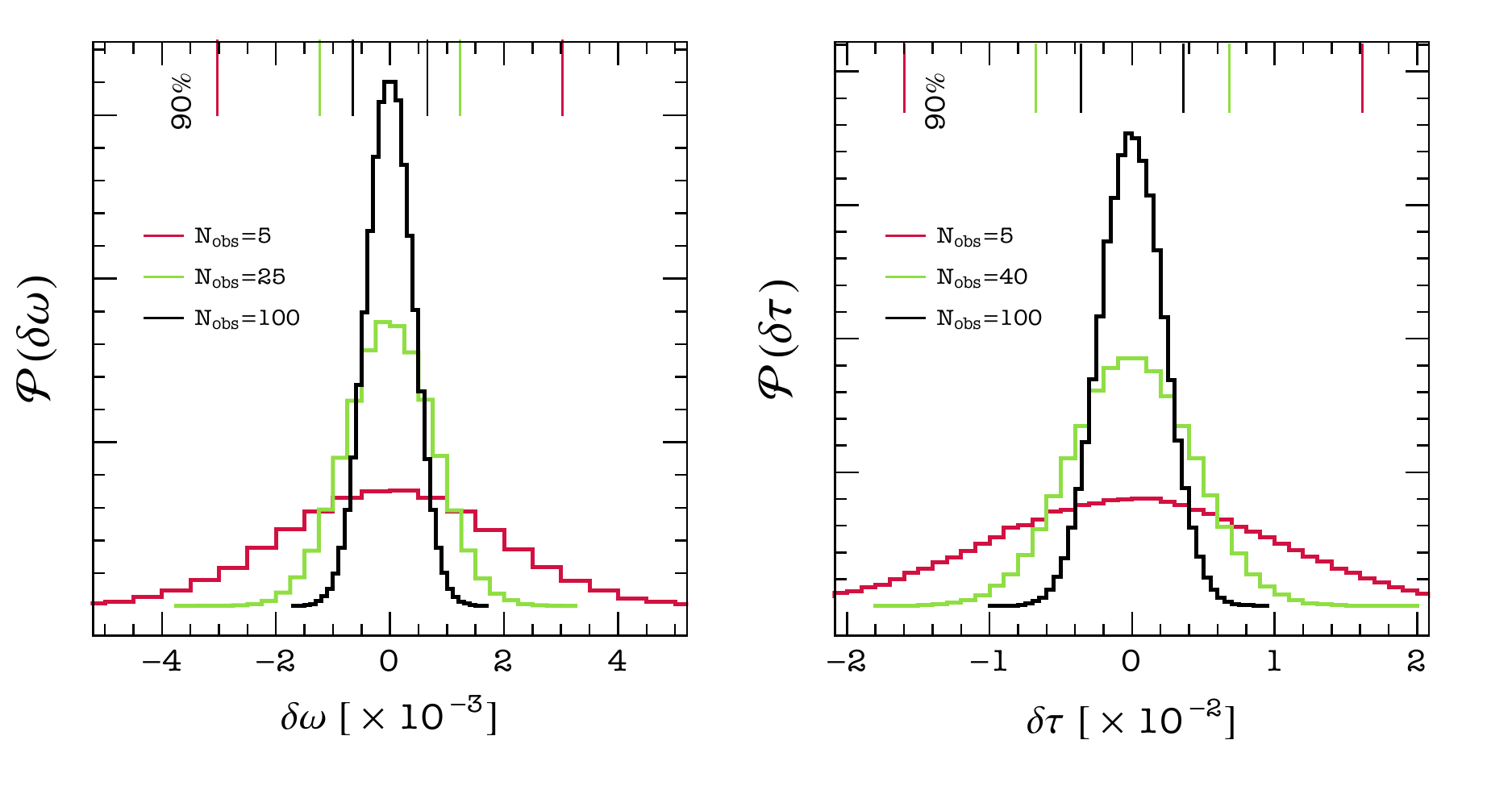}
\caption{Posterior probability distributions  
for GR deviations within the QNMs spectrum of 
rotating BHs observed by LGWA, as a function 
of the number of observed events. We assume 
only the fundamental $\ell=m=2$ mode is detected, 
and that the SNR of each event is fixed to 100.
Vertical bars identify 90\% confidence intervals. 
Time is measured in units $GM/c^3$.}
\label{fig:ppetest2}
\end{figure}

LGWA will also be able to test a fundamental pillar of GR, 
i.e. the uniqueness of the Kerr nature \cite{Hansen:1974zz}, 
through observations BH quasi normal modes (QNMs). The 
latter are completely determined, in GR, by the mass and 
spin angular momentum of the ringing object, while solutions 
beyond GR may feature a non trivial dependence on extra 
parameters \cite{Berti:2015itd}. Any change in the QNM 
frequencies $\omega$ and damping times $\tau$ of rotating 
BHs can be parametrized as: 
\begin{equation}
\omega=\omega^\textnormal{Kerr}+\delta\omega\ ,\quad
\tau=\tau^\textnormal{Kerr}+\delta\tau\ ,
\end{equation}
where $\delta\omega$ and $\delta\tau$ are GR deviations 
that can be constrained by actual GW signals
\cite{Gossan:2011ha,Carullo:2018sfu,Meidam:2014jpa,Maselli:2019mjd}.
We explore here the possibility of LGWA to perform null 
test of GR, i.e. assuming $\delta \omega=\delta\tau=0$, 
by constraining the QNM shift parameters using multiple 
observations of the BH fundamental $\ell=m=2$ mode 
\cite{Maselli:2019mjd}. Figure \ref{fig:ppetest2} shows 
the probability distribution inferred on $\delta\omega$ 
and $\delta\tau$, using a Bayesian approach \cite{Maselli:2019mjd}, 
with an increasing number of detected modes, from 
binaries with component masses and spins uniformly drawn 
between $(10^6,10^7)M_\odot$, and $(-1,1)$, respectively. 
For sake of simplicity we assume that the SNR of each mode 
is equal to 100. LGWA can potentially narrow GR deviations 
with good accuracy, on both frequencies and damping  times, 
with one hundred of events leading to constrain 
$\vert \delta\omega \vert\lesssim 7\cdot10^{-4}$ and 
$\vert \delta\tau \vert \lesssim 4\cdot10^{-3}$ at 90\% 
confidence level. Here, time is expressed in units $GM/c^3$, with $M$ being the mass of the black-hole remnant.

A unique ability of LGWA as a single detector to test GR originates from the 
Moon's spherical shape and the fact that seismometers can be distributed, in principle, over its entire surface. This idea was first published for the concept of a spherical, resonant detector \cite{WaPa1976,BiEA1996}, and it requires an estimation of polarization content of a GW field. General relativity predicts that GWs can only exist as tensor mode, but other metric theories might give rise to scalar and vector modes. In total, 6 independent polarization modes can exist in metric theories. Assessing whether a GW signal is consistent with a pure tensor mode constitutes a GR test. As demonstrated for the Virgo/LIGO detector network, the test can be done on stochastic GW backgrounds \cite{AbEA2018b} or on individual signals as for, e.g., the BNS merger GW170817 \cite{AbEA2019b}. In all cases so far, no evidence for scalar and vector modes was found.

\subsection{Lunar science}
The Moon is a complex differentiated planetary object and the only body besides the Earth on
which extensive seismological experiments have been carried out with success. In fact, Apollo
missions installed a network of four 3-component, long period seismometers on the Moon (two other stations were short-lived), which revealed more than 12,500 seismic events over a 9 year period \cite{NaEA1981,KhEA2013}. Even if our knowledge of the Moon has been significantly increased with Apollo missions, much remains to be explored and discovered, both regarding the origin (and history of the Earth-Moon system) and the internal structure. The absence of plate tectonics made it possible that well-defined accretion and geological evolution records are preserved, unlike on Earth.

\begin{figure}[ht]
\centering
\includegraphics[width=1\columnwidth,height=0.6\columnwidth]{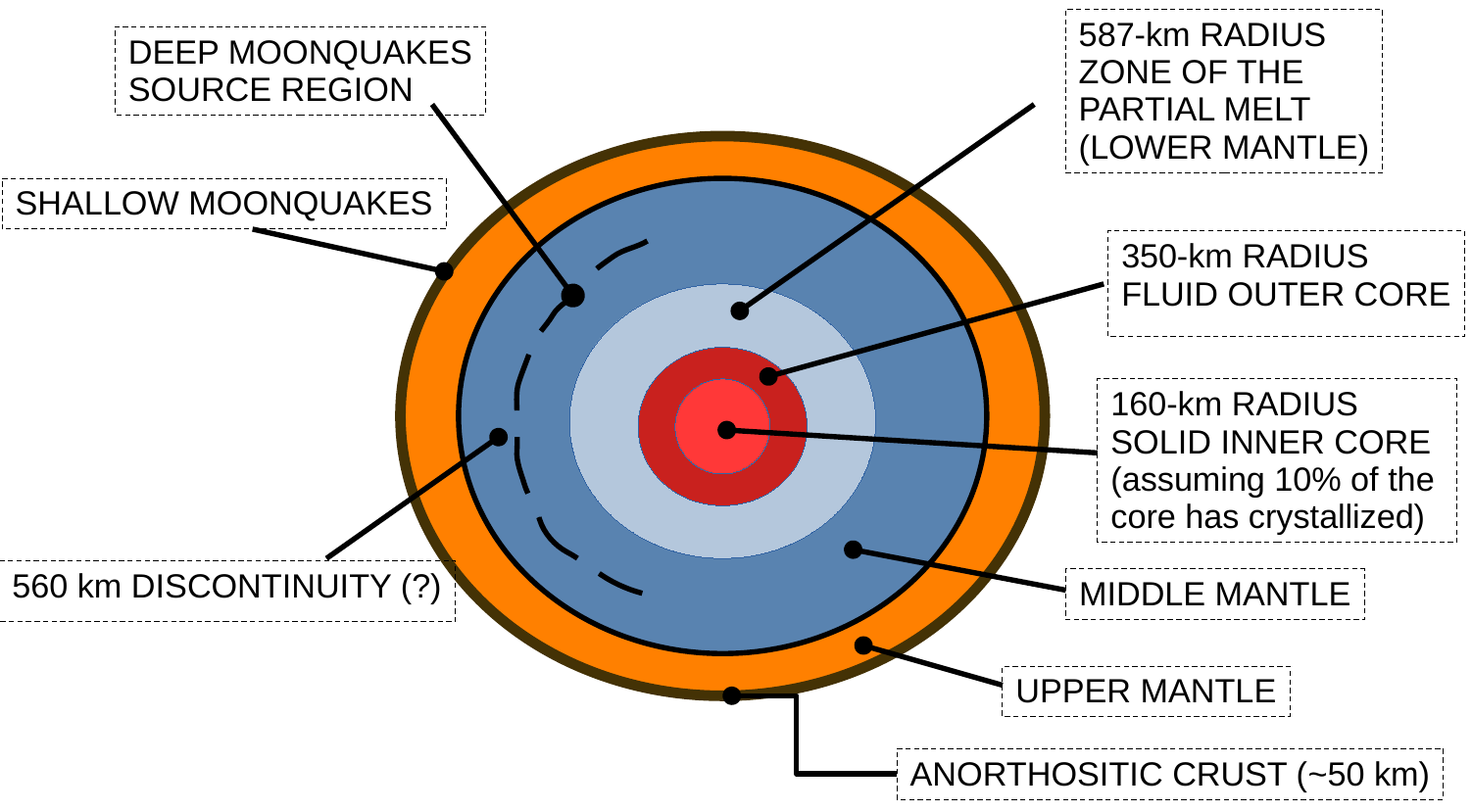}
\caption{Schematic of the internal structure of the Moon. The Moon's radius is 1737\,km.}
\label{fig:lunarsketch}
\end{figure}

Several models have been proposed in order to explain the origin of the Moon: co-accretion (e.g., \cite{Thomson1864}), fission (e.g., \cite{Durisen1984}) and capture (e.g., \cite{Urey1952}), but all these scenarios suffer from serious flaws. The current accepted model is the so-called "Giant impact" which implies the collision of the proto-Earth with a Mars-size differentiated object in the early solar system \cite{Canup2004}. This model would be able to explain why, for example, the Moon has not a large iron core and why it has exactly the same oxygen isotope composition as the Earth. The interior structure is probably differentiated into a crust, a mantle and a core (see figure \ref{fig:lunarsketch}), even if the degree of differentiation is low as attested by its non-dimensionalized moment of inertia, i.e., 0.393 \cite{Williams2014}. As a comparison, the Earth has a non-dimensionalized moment of inertia of 0.33 \cite{Williams1994}.

\begin{figure*}[ht]
\centering
\includegraphics[width=2\columnwidth]{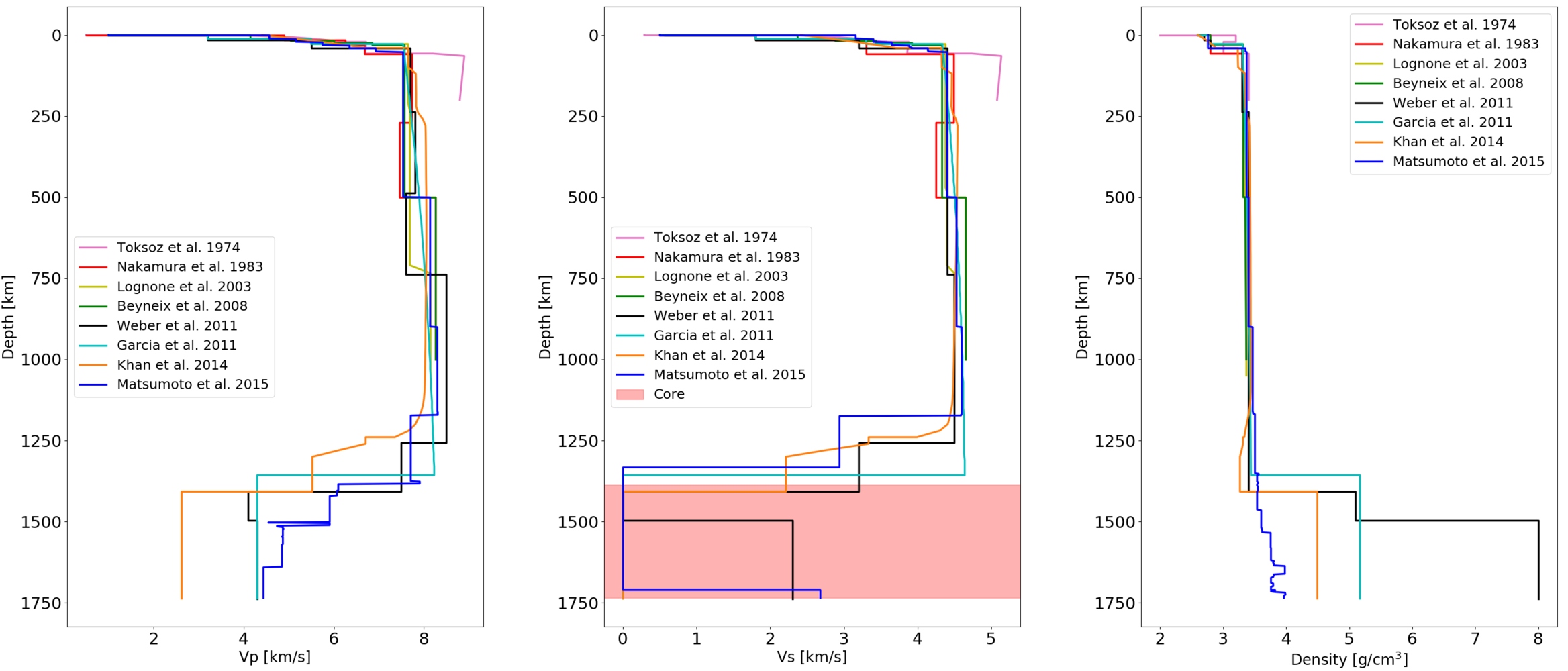}
\caption{Comparison of several seismic velocity models (left and middle plots for compressional-wave speed V$_{rm p}$ and shear-wave speed V$_{\rm s}$), and density (right). In the middle plot, we identify also the core, assuming a size of $\simeq$ 350\,km \cite{Wieczorek2006}. Data for these plots are taken from Garcia et al.~\cite{GaEA2019}. These models are important for the calculation of normal modes and their response to GWs. The references in the plot legends are \cite{Toksoz1974,Nakamura1983,Lognonne2003,Beyneix2006,Weber2011,Garcia2011,Khan2014,Matsumoto2015}.}
\label{fig:lunarstruct}
\end{figure*}

The crust has a thickness of about 50\,km and an anorthositic composition, containing elements like O, Si, Fe, Al and also long-lived radiogenic elements \cite{GaEA2019}. The mean density is about 2500\,kg\,m$^{-3}$ with a porosity of 12$\%$, which probably increases in a significant way with depth \cite{Besserer2014}. The composition of the mantle is predominantly made of olivine, orthopyroxene and clinopyroxene \cite{Toksoz1974,HJ1987}, more iron rich than on Earth. The uppermost mantle has a density of about 3200\,kg\,m$^{-3}$ and a porosity of $6\%$, with a thermal gradient of 0.5-0.6\,$^\circ$C/km \cite{GaEA2019}. Seismic data suggest a discontinuity at about 560\,km from the surface. At a depth of about 1150\,km, instead, seismic waves are strongly attenuated in a region which is partially melted and probably represents the core-mantle boundary \cite{Nakamura2005,Weber2011}. The core is small ($\simeq$ 350\,km - \cite{Wieczorek2006}), liquid in the external and solid in the internal part: this configuration is required to explain the lunar laser ranging (LLR) measurements of the Moon's pole rotation (e.g. \cite{Williams2001}). However, there are several estimates of the size and density (composition) of the core derived from geophysical data and modeling: please see the Tab.~1 of Garcia et al. \cite{GaEA2019} for a complete review. Even if the composition of the core is not well constrained, it could be composed mostly of metallic iron alloy and a small quantity of sulfur and nickel \cite{GaEA2019}. As discussed before, the key information about the internal structure comes from seismic studies. Several lunar seismic models exist: see figure \ref{fig:lunarstruct}. These models have greatly increased our knowledge about the deep lunar interior, for example, the mantle density structure, but they have substantial difficulty to well define the core size, since the core is very small.

Seismicity on the Moon can be classified in three main categories: the deep moonquakes, occurring in the so-called "nests" (at a depth between 700 -- 1100\,km), with typical magnitudes of around 2, correlated with the lunar tides; the shallow moonquakes are stronger (up to magnitudes of 5.5 on the Richter scale) than the deep moonquakes, but relatively rare and occurring at depths between 50 -- 200\,km and not connected with geological or geographic features (e.g. \cite{Nakamura1979}). Finally, moonquakes can also be caused by meteoroid impacts.

\subsection{Synergy with other detectors and multiband observations}
\label{sec:complement}
Synergy emphasizes the mutual benefit from combined observations with other GW detectors or with electromagnetic facilities to improve their combined science case. The science case outlined in the previous sections contained elements of synergy focused mainly on multi-messenger observations. In the following, we briefly discuss the synergy between GW detectors.

If a few GW detectors are sensitive in the same band, then a well-known example of synergy is the localization of a GW source in the sky \cite{WeCh2010,AbEA2017d}. The source localization of the BNS merger GW170817 was essential for the ensuing multi-messenger campaign \cite{AbEA2017e}. Estimation of other parameters like distance and inclination angle of a compact binary can also improve when data from several detectors are combined. For LGWA, it is hard to make any predictions whether it will observe with another detector with overlapping observation band, but the space-borne detector LISA is a possible example (provided that LGWA can be realized on a similar timeline). Taking into account that LGWA has interesting sky-localization capabilities due to its more rapid rotation (27.3 days compared to 1 year for LISA), it can be expected that it would significantly improve parameter estimation of some GW signals compared to LISA alone. 

If GW detectors are sensitive in different frequency bands, then joint multiband observations can be carried out \cite{Ses2016}. Signals from compact binaries evolve in frequency and can sweep over various bands. For example, solar-mass compact binaries or intermediate-mass BBHs can inspiral from the LGWA band to the band of ground-based GW detectors within a few years determined by equation (\ref{eq:insptime}). This can lead to improved parameter-estimation accuracy \cite{IsEA2018,GrHa2020}. We have seen in section \ref{sec:fundamental} as well that multiband observations can lead to improved tests of GR. Concerning LGWA, very interesting is the case of inspiraling BNS. If first observed at frequencies above 0.1\,Hz, these would have merger times short enough --- up to a few years --- to make multiband observations with ground-based detectors possible. LGWA could issue an early warning to ground-based detectors of such highly interesting events \cite{ChEA2018}. 

\section{Conclusions}
\label{sec:conclusions}
In response to the ESA Call for Ideas, ``Exploring the Moon with a large European lander'', a concept for a GW detector on the Moon --- the Lunar GW Antenna (LGWA) --- was submitted \cite{HaEA2020b}. It is based on the early idea of Joseph Weber to monitor surface vibrations of the Moon caused by GWs. A near-term opportunity to explore lunar environmental parameters of the Moon important to LGWA is provided by NASA's Moon exploration program Artemis, which includes human landings starting with Artemis III and several robotic landings. An Artemis III science white paper was submitted to propose the deployment of a lunar geophysical station in one of the permanent shadows of the south pole, which we identified as a potentially very attractive location of a future LGWA station deployment \cite{HaEA2020c}. 

In this paper, we have described the LGWA concept in greater detail. We have shown that LGWA would have a broad science case as an independent GW detector and in synergy with other detectors. The science can be enriched by electromagnetic observations allowing a multi-messenger astronomy of sources such as inspirals and mergers of white-dwarf binaries, supermassive, massive and intermediate-mass black-hole binaries, and potentially, although less likely, spinning neutron stars. Furthermore, LGWA would contribute to the geophysical exploration of the Moon, and vice versa, GW analyses of LGWA data crucially rely on the availability of accurate models of Moon's internal structure.

Central to the concept is a seismometer whose sensitivity must approximately match the displacement sensitivity of the LISA Pathfinder mission, which was, however, demonstrated under conditions of near free fall. Achieving similar performance under the influence of Moon's gravity, which requires a suspension mechanism of the seismometer's test mass, is a nontrivial feat. 

Operating seismometers of such sensitivity would require an extremely low level of seismic disturbances, which cannot be found in natural, terrestrial environments. Since the instrument performance needs to be demonstrated ahead of a deployment on the Moon, the development of a seismic platform stabilized actively using high-end seismic sensors of ground displacement and rotation is necessary. Such platforms would ideally be located in underground environments to profit from a straight-forward reduction of the seismic input.

An opportunity for the required technology developments for LGWA is the strong overlap with technologies required for future, terrestrial GW detectors like Einstein Telescope and Cosmic Explorer. Platforms with actively suppressed seismic motion have already been realized as part of a seismic-isolation system for the Advanced LIGO detectors \cite{MaEA2014,MaEA2015}, but their performance needs to be improved significantly for LGWA especially to extend the noise suppression towards mHz frequencies. Also, increasing the sensitivity of seismometers beyond what is currently available commercially or in research laboratories could equally lead to improved performance of seismic isolation systems in future GW detectors enabling the most optimistic scenarios of low-frequency, terrestrial GW observations. In this context, development of LGWA technologies has already started.

\appendix
\section{Meteoroid background}
\label{app:meteoroid}
For a simplified evaluation of seismic waves produced by meteoroid impacts, we assume that the meteoroid flux is stationary and isotropic. We also assume that the impact is in normal direction to the surface, which means that we can focus on the radial displacement of spheroidal modes. It should be straight-forward to extend the calculation to arbitrary angles of incidence, so that meteoroid impacts can excite radial and transverse displacement of spheroidal and toroidal modes.

Each impact event can be represented as a surface point force at angular coordinates $\theta_0,\,\phi_0$. The normal surface displacement at $\theta,\phi$ is determined by a Green's function, which can be written as the following product \cite{Ben1983}
\begin{equation}
\begin{split}
    _nu^{\rm S}_{lm}(\theta,\phi;\omega)&=\;_nG_{lm}(\theta,\phi|\theta_0,\phi_0;\omega) f(\theta_0,\phi_0;\omega)\\
    &=(Y_l^m(\theta,\phi))^*\,Y_l^m(\theta_0,\phi_0)\frac{y_{1n}^2(R)}{_n\Lambda_{lm}}\\
    &\qquad\cdot \frac{f(\omega)}{\omega_n^2-\omega^2+\ii\omega_n^2/Q_n},
\end{split}
\label{eq:metresponse}
\end{equation}
where $R=1.7\cdot10^6\,$m is the radius of the Moon, $y_{1n}$ is a radial function characterizing radial displacements of spheroidal normal modes, $f(\omega)$ is the Fourier domain amplitude of the radial force exerted by the impact, and $Y_l^m(\theta,\phi)$ are the scalar surface spherical harmonics. The normalization factor
\begin{equation}
    _n\Lambda_{lm}=\int\limits_0^R\dd r\,r^2\rho_0(r)(y_{1n}^2(r)+l(l+1)y_{3n}^2(r))
\end{equation}
implies that the Moon is modeled as a laterally homogeneous body (its density $\rho_0$ only depends on the radius). Such models have been used with great success in normal-mode analyses of Earth vibrations. The radial function $y_{3n}$ appearing inside this integral characterizes the transverse displacement of spheroidal modes. This definition of the radial normalization factor is consistent with spherical harmonics in orthonormal normalization:
\begin{equation}
    \int\dd\Omega\,(Y_l^m(\theta,\phi))^*Y_{l'}^{m'}(\theta,\phi)=\delta_{ll'}\delta_{mm'}.
\end{equation}
A different normalization convention is used by Ben-Menahem, which means that numerical factors are not the same when comparing with his work \cite{Ben1983}. We now proceed by calculating the power spectral density of background noise $_nS_{lm}(\omega)$ integrating over the Moon's surface with respect to angles $\theta_0,\,\phi_0$ assuming an isotropic meteoroid shower. In the following, we can omit the angular dependence $Y_l^m(\theta,\phi)$ of the excited normal modes, since we only need the normal-mode amplitudes. We then obtain
\begin{equation}
    _nS_{lm}(\omega) = \frac{S_p(\omega)}{(\omega_n^2-\omega^2)^2+\omega_n^4/Q_n^2}\frac{R^4y_{1n}^4(R)}{_n\Lambda_{lm}^2},
\end{equation}
where $S_p(\omega)$ is the spectral density of effective pressure fluctuations acting on the Moon's surface isotropically due to a steady flux of meteoroids, which also depends on the formation of ejecta, elasticity properties of the impact, and the transfer efficiency of impact energy to the seismic field (typically a small fraction of the kinetic energy of a meteoroid) \cite{LoEA2009}. The last equation  has the on-resonance form
\begin{equation}
    _nS_{lm}(\omega_n) = \frac{Q_n^2}{\omega_n^4}S_p(\omega_n)\frac{R^4y_{1n}^4(R)}{_n\Lambda_{lm}^2}.
\end{equation}
The term $y_{1n}^2(R)/_n\Lambda_{2m}$ depends on the Moon's internal structure and is of the order of $0.2/M$ for lower-order quadrupole modes, where $M=7.3\cdot 10^{22}\,$kg is the mass of the Moon, so that we can approximate the last equation as
\begin{equation}
    _nS_{lm}(\omega_n) \approx 0.04\frac{Q_n^2}{\omega_n^4}\frac{R^4S_p(\omega_n)}{M^2}.
    \label{eq:noise}
\end{equation}
The spectral density of pressure fluctuations is related to pressure rms by $S_p(\omega)=\langle p^2\rangle \tau/(1+(\omega \tau)^2)\approx \langle p^2\rangle \tau$ below 1\,Hz \cite{TGR2017}, where $\tau$ is the average duration of meteoroid impacts henceforth assumed to be $\tau=0.1\,$s neglecting complications like the formation of ejecta.

The effective pressure $\langle p\rangle$ produced by meteoroids with flux $F$ and momentum $mv$ is given by
\begin{equation}
    \langle p\rangle = \alpha\langle F mv\rangle
\end{equation}
where $\alpha$ is the efficiency of momentum transfer to the seismic field. For our order-of-magnitude estimate, we go with the approximation $\langle p^2\rangle = (\alpha F_0m_0v_0)^2$, where we use $F_0=10^{-5}\,\rm m^{-2}s^{-1}$ for the flux and $m_0v_0=10^{-3}\,$kg\,m/s for the average meteoroid momentum \cite{GHS2011}, and $\alpha=0.01$, which seems to be a generous overestimation of the impact-energy-to-seismic conversion efficiency  \cite{LoEA2009,QuEA2019}. Note that this expression is independent of the \emph{dynamical} time scale $\tau$ of the impact since we are interested in the average force or pressure, which means that the relevant time scale $\Delta t$ of momentum transfer $\Delta P$ is given by the average time between impacts, which is contained in the meteoroid flux.

It is possible that the total (undirected) momentum transfer onto the Moon is dominated by rare impacts of larger meteoroids, where rare means less frequent than once per $Q_n/f_n$, which is about once per day for the lower order quadrupole modes. With $4\pi R^2\times 1\,\rm day=3\cdot 10^{18}\,m^2s$, we estimate that impacts of meteoroids with mass greater than 1\,kg occur less than once per day \cite{GHS2011}. These events need to be treated individually, and cannot be considered part of the stationary background noise anymore. As for any other larger seismic disturbance, one needs to exclude the stretch of data containing the large impact event from GW analyses, or attempt a subtraction of the seismic signal from the seismometer data.

Inserting the expression of the surface pressure from meteoroid impacts into equation (\ref{eq:noise}), we obtain
\begin{equation}
    \sqrt{_nS_{lm}(\omega_n)} \approx 0.005\frac{Q_n}{f_n^2}\frac{\alpha R^2F_0m_0v_0}{M}\sqrt{\tau}
\end{equation}
Confronting this equation with the on-resonance normal-mode response to GWs, we can say that the effective strain noise in mode $n$ from meteoroid impacts is
\begin{equation}
\begin{split}
    h^n_{\rm met}&=0.01\frac{\alpha R^2F_0m_0v_0}{L_nf_n^2M}\sqrt{\tau}\\
    & \approx 2\times10^{-23}\,\rm Hz^{-1/2},
\end{split}
\label{eq:metstrain}
\end{equation}
where the numerical value depends weakly on the order of normal modes below about 10\,mHz according to our simplified response model. The value lies well below the LGWA strain-noise target of $10^{-20}\,\rm Hz^{-1/2}$. For a full frequency-dependent expression, one needs to include off-resonance response of normal modes as present in equation (\ref{eq:metresponse}) and sum over all modes. Also, the full GW response is required when referring the meteoroid background to effective GW strain noise as done on resonance in equation (\ref{eq:metstrain}). Finally, one needs to keep in mind that the numerical factor in the last equation depends on the order of the normal mode. Still, the on-resonance meteoroid background in the mHz band can be approximated by the formula in equation (\ref{eq:metstrain}).

\section*{Acknowledgements}
AM acknowledges support from the Amaldi Research Center, funded by the MIUR program "Dipartimento di Eccellenza", CUP: B81I18001170001. JH acknowledges support from PRIN-MIUR 2017 (grant 2017SYRTCN). VB, RDC, PS and RS acknowledge partial financial contribution from the agreement ASI-INAF n.c2017-14-H.O. EC, EB, MB, MdV, AG, AP,  PDA, EP acknowledge support from PRIN-MIUR 2017 (grant 20179ZF5KS).

\bibliographystyle{apsrev} 
\bibliography{references}

\end{document}